\pdfoutput=1
\documentclass[11pt]{article}
\usepackage[utf8]{inputenc}
\usepackage{amsmath}
\usepackage{amssymb}
\usepackage{comment}
\usepackage{float}
\usepackage{cite}
\usepackage{caption}
\usepackage{xcolor}
\usepackage{tikz}
\usetikzlibrary{arrows.meta,calc,shadows}
\definecolor{purple}{rgb}{0.4,.2,0.7}
\definecolor{cEps}{rgb}{0.122,0.467,0.706}  %
\definecolor{cSR}{rgb}{0.80,0.12,0.18}      %
\definecolor{cRes}{rgb}{0.48,0.24,0.68}     %
\definecolor{cHT}{rgb}{0.12,0.52,0.38}      %
\definecolor{cNRG}{rgb}{0.85,0.39,0.08}     %
\definecolor{cMC}{gray}{0.12}              %
\usepackage[margin = 2cm]{geometry}
\usepackage{graphicx}
    \graphicspath{{figures/}}
\usepackage[hyperfootnotes = false, colorlinks = true, linkcolor = blue, citecolor = purple, urlcolor = blue]{hyperref}
\usepackage{subcaption}
\usepackage{multicol,multirow}
\usepackage{booktabs}
\usepackage{makecell}

\parskip=\medskipamount

	\usepackage{physics}
	\newcommand{\ii}{\mathrm{i}} %

\newcommand{\SM}{\href{https://github.com/Canonical111/LRI_public/blob/main/Supplementary\%20Material.pdf}{Supplemental Material}}

\definecolor{ptnew}{rgb}{0,0.25,0.7}
\definecolor{ptnum}{rgb}{0.75,0.1,0.1}
\newcommand{\ptPub}[1]{$#1$}
\newcommand{\ptClosed}[1]{\textcolor{ptnew}{${#1}^{\ast}$}}
\newcommand{\ptExact}[1]{\textcolor{ptnew}{${#1}^{\ddagger}$}}
\newcommand{\ptConj}[1]{\textcolor{ptnew}{${#1}^{\text{\S}}$}}
\newcommand{\ptNum}[1]{\textcolor{ptnum}{${#1}^{\dagger}$}}

\newcommand{\DE}{\Delta_{\mathcal E}}
\newcommand{\Ds}{\Delta_\sigma}

\newcommand{\CfourOne}{45.42488010356}       %
\newcommand{\CfiveOne}{-731.13947662454}     %
\newcommand{\PTorderOneDword}{eight}       %
\newcommand{\PTorderLitword}{three}        %
\newcommand{\CsixOne}{14331.68703275805}     %
\newcommand{\CsevenOne}{-322550\pm40} %
\newcommand{\CeightOne}{(8.060\pm0.017)\times10^{6}} %
\newcommand{\CsixTwo}{362.3665\pm0.0032} %
\newcommand{\CsevenTwo}{-3876.1\pm2.1}       %
\newcommand{\CsixThree}{33.5754\pm0.0005}    %
\newcommand{\CsevenThree}{-220.35\pm0.13}    %
\newcommand{\CtwoOne}{0.657131263981}        %
\newcommand{\CthreeOne}{-3.719856322902}     %
\newcommand{\CtwoTwo}{0.308065413582}        %
\newcommand{\CthreeTwo}{-0.874850012653}     %
\newcommand{\CtwoThree}{0.181221785406}      %
\newcommand{\CthreeThree}{-0.353581520433}   %
\newcommand{\CfourTwo}{5.048868431}          %
\newcommand{\CfiveTwo}{-38.833645896}        %
\newcommand{\CfourThree}{1.228063545}        %
\newcommand{\CfiveThree}{-5.850351421}       %
\newcommand{\CeightTwo}{45990\pm33}          %
\newcommand{\CeightThree}{1603\pm4} %
\newcommand{\PTdThreeDigits}{50}             %
\newcommand{\SRorderword}{eight}           %
\newcommand{\SRorderLitword}{two}          %
\newcommand{\SRCthree}{5.72737}  %
\newcommand{\SRCfour}{-12.1962}  %
\newcommand{\SRCfive}{29.2378}  %
\newcommand{\SRCsix}{-58.6657}  %
\newcommand{\SRCseven}{51.0690}  %
\newcommand{\SRCeight}{247.381}  %
\newcommand{\SRCeightMag}{247.38\pm0.13} %
\newcommand{\SRtwoCtwo}{-4.06627}  %
\newcommand{\SRtwoCtwoMC}{-4.075\pm0.019} %
\newcommand{\SRintegClosedDev}{7\times10^{-26}} %
\newcommand{\SRCeightMagDiff}{2.5\times10^{-5}} %
\newcommand{\SRthreeCone}{0.2732}        %
\newcommand{\SRthreeConeRange}{[0.2690,\,0.2760]} %
\newcommand{\SRthreeConeBRRZ}{0.27(4)}   %

\newcommand{\MCnAcceptedword}{nine}    %

\newcommand{\MCthreeD}{1.47314(118)(230)} %
\newcommand{\MConeL}{2^{10}\text{ to }2^{19}} %
\newcommand{\MCtwoL}{64\text{ to }1024}      %
\newcommand{\MCthreeL}{8\text{ to }128}      %

\newcommand{\HTpLamThermal}{118}        %
\newcommand{\HTpOneLexact}{2000}        %
\newcommand{\NOneCfiveLmax}{500}       %
\newcommand{\HTpLamCeight}{92}          %
\newcommand{\HTpCtwoOne}{0.657131263984(10)}     %
\newcommand{\HTpCtwoOneDev}{4\times10^{-12}}   %
\newcommand{\HTpCthreeOne}{-3.7198563(10)}       %
\newcommand{\HTpCthreeOneDev}{2\times10^{-8}}   %
\newcommand{\HTpCfourOne}{45.4249(9)}           %
\newcommand{\HTpCfourOneDev}{1\times10^{-6}}    %
\newcommand{\HTpCfiveOne}{-731.13(10)}         %
\newcommand{\HTpCfiveOneDev}{2\times10^{-5}}    %
\newcommand{\HTpCsixOne}{14\,320\text{--}14\,342} %
\newcommand{\HTpCsixOneDev}{8\times10^{-4}}     %
\newcommand{\HTpCsevenOne}{-322\,890\text{ to }{-322\,230}} %
\newcommand{\HTpCsevenOneDev}{1\times10^{-3}}   %
\newcommand{\HTpCeightOne}{8.0652\times10^{6}}  %
\newcommand{\HTpCeightOneDev}{7\times10^{-4}}   %
\newcommand{\HTpCeightOneUncal}{8.045\text{--}8.073}    %
\newcommand{\HTpCtwoTwo}{0.30806541358}         %
\newcommand{\HTpCtwoTwoDigits}{35}              %
\newcommand{\HTpCthreeTwo}{-0.8748500(1)}       %
\newcommand{\HTpCthreeTwoDev}{1\times10^{-8}}   %
\newcommand{\HTpCfourTwo}{5.0488682(10)}        %
\newcommand{\HTpCfourTwoDev}{5\times10^{-8}}    %
\newcommand{\HTpCfiveTwo}{-38.835(13)}         %
\newcommand{\HTpCfiveTwoDev}{5\times10^{-5}}    %
\newcommand{\HTpTwoLowDev}{10^{-7}}             %

\newcommand{\HTnpNmax}{40}                 %
\newcommand{\HTnpDensK}{16}                %
\newcommand{\HTnpBasisErr}{10^{-6}} %

\newcommand{\HTnpRawViol}{0.06\text{ to }0.25} %
\newcommand{\HTnpWinReduce}{10\text{ to }40}   %
\newcommand{\HTnpResSig}{0.01}            %
\newcommand{\HTnpResE}{0.05}              %
\newcommand{\HTnpViolN}{16\text{ to }28}  %
\newcommand{\HTnpIdentN}{28}             %

\newcommand{\HTnpEpsDiff}{-4\times10^{-8}} %
\newcommand{\HTnpNfortyFifty}{0.620671}          %
\newcommand{\HTnpNfortyTwentyFive}{0.710577}     %
\newcommand{\HTnpNfortyTen}{0.809111}            %
\newcommand{\HTnpNfortyOne}{0.969338}            %
\newcommand{\HTnpDEtwoFortyNine}{0.49934358(9)} %
\newcommand{\HTnpDEtwoForty}{0.4942198(26)} %
\newcommand{\HTnpDEtwoTwentyFive}{0.488080(5)}     %
\newcommand{\HTnpDEtwoHundred}{0.483024(5)}        %
\newcommand{\HTnpDEoneSeventyFive}{0.483909(15)}   %
\newcommand{\HTnpDEoneFifty}{0.491209(33)}         %
\newcommand{\HTnpDEoneTwentyFive}{0.50612(4)}      %
\newcommand{\HTnpDEoneHundred}{0.53048(12)}        %
\newcommand{\HTnpDEseventyFive}{0.56733(20)(-50)}  %
\newcommand{\HTnpDEfiftyFOneTwo}{0.6220\text{--}0.6224} %
\newcommand{\HTnpDEfiftyFthree}{0.618}             %
\newcommand{\HTnpTrust}{10^{-4}}                 %

\newcommand{\HTnpDtwoNmax}{24}             %

\newcommand{\HTnpDthreeNmax}{20}           %
\newcommand{\HTnpDthreeSR}{1.41924}        %
\newcommand{\HTnpRawShift}{(0.7\text{--}2.3)\times10^{-3}} %

\newcommand{\NRGnPoints}{13}            %

\newcommand{\NRGDETwoFortyNine}{[0.5080]} %
\newcommand{\NRGDETwoForty}{[0.4989]} %
\newcommand{\NRGDETwoTwentyFive}{[0.4897]} %
\newcommand{\NRGDETwoHundred}{0.4830(7)} %
\newcommand{\NRGDEOneSeventyFive}{0.48390(35)} %
\newcommand{\NRGDEOneFifty}{0.49111(11)} %
\newcommand{\NRGDEOneTwentyFive}{0.50567(5)} %
\newcommand{\NRGDEOneHundred}{0.529608(33)} %
\newcommand{\NRGDESeventyFive}{0.566260(30)} %
\newcommand{\NRGDEFifty}{0.622115(27)} %
\newcommand{\NRGDETwentyFive}{0.713422(29)} %
\newcommand{\NRGDETen}{0.80975(5)} %
\newcommand{\NRGDEOne}{0.93742(11)} %

\newcommand{\RSKI}{6.2466\times10^{-3}}       %
\newcommand{\RSgffOnlyEnd}{0.567}             %
\newcommand{\RSoneDmin}{0.4828}               %
\newcommand{\RSoneDminAt}{0.192}              %
\newcommand{\RSoneDenv}{4.7\times10^{-3}}     %
\newcommand{\RStwoDenv}{5.2\times10^{-3}}     %
\newcommand{\RSthreeDenv}{1.5\times10^{-4}}   %
\newcommand{\RStwoDendpoint}{7.5\times10^{-3}} %
\newcommand{\RSgffOnlyTwo}{1.0049}            %
\newcommand{\RSgffOnlyThree}{1.4161}          %
\newcommand{\RSgffDevTwo}{5.2\times10^{-3}}   %
\newcommand{\RScbThreeTwo}{0.087}             %

\newcommand{\CMPhtCB}{5\times10^{-4}}       %
\newcommand{\CMPnrgCB}{7.6\times10^{-4}}    %
\newcommand{\CMPmcCB}{3.2\times10^{-3}}     %
\newcommand{\CMPmcGauss}{0.006\text{--}0.010} %
\newcommand{\CMPtwoUfour}{0.0015}           %
\newcommand{\CMPtwoUfourEnds}{0.005\text{--}0.02} %
\newcommand{\CMPthreeMC}{1.0\times10^{-3}}  %
\newcommand{\CMPthreeMCsr}{0.022\text{--}0.033} %

\newcommand{\CMPkfourKeight}{1.3\times10^{-3}} %

\newcommand{\SRtwoCtwoHT}{-4.0670(24)}        %
\newcommand{\SRtwoCtwoHTDev}{2\times10^{-4}}  %

\newcommand{\HTpCtwoThree}{0.181221785405554}   %
\newcommand{\HTpCtwoThreeDigits}{15}
\newcommand{\HTpCthreeThree}{-0.353581479(20)}  %
\newcommand{\HTpCthreeThreeDev}{1\times10^{-7}} %
\newcommand{\HTpCfourThree}{1.228070(21)}       %
\newcommand{\HTpCfourThreeDev}{5\times10^{-6}}  %
\newcommand{\HTpCfiveThree}{-6.0\pm0.3}         %
\newcommand{\HTpCfiveThreeDev}{3\times10^{-2}} %
\newcommand{\HTpThreeLfour}{600}                %
\newcommand{\HTpThreeLfive}{40}                 %
\newcommand{\RSnineOnePct}{3.5\times10^{-4}}   %
\newcommand{\RSnineCeightAmp}{nine}            %
\newcommand{\RSnineImplied}{-2.190\times10^{8}\text{ and }{-2.178}\times10^{8}} %
\newcommand{\RSnineSpread}{0.5}                %

\begin{document}
\renewcommand*{\thefootnote}{\fnsymbol{footnote}}

\thispagestyle{empty}
\begin{center}
    ~\vspace{5mm}

{\LARGE \bf The long-range Ising CFT: \\[2mm] perturbative and nonperturbative approaches}

   \vspace{0.5in}
\begingroup\large
   {\bf Zechuan Zheng\footnote{zechuan.zheng.phy@gmail.com}}
\par\endgroup
    \vspace{0.5in}

\begingroup\sffamily\footnotesize
    Perimeter Institute for Theoretical Physics, Waterloo, ON N2L 2Y5, Canada
 \par\endgroup
    \vspace{0.5in}

\end{center}

\vspace{0.5in}

\begin{abstract}
The long-range Ising model is the prototype of criticality with power-law interactions. In each dimension its interacting critical points form a one-parameter family of nonlocal conformal field theories, weakly coupled at both ends but not in between. We study this family in one, two and three dimensions from several independent directions. We extend the expansion about the generalized free field from three to eight loops and expand about the short-range end. We show that Hamiltonian truncation applies to this nonlocal theory once a truncation artifact is corrected. Where they can be compared, the resummed expansions, Hamiltonian truncation, the numerical renormalization group and Monte Carlo simulations agree on the dimension of the leading $\mathbb Z_2$-even scalar at the sub-percent level, as universality requires.
\end{abstract}

\vspace{0.5cm}

\thispagestyle{empty}

\newpage

\setcounter{tocdepth}{3}

\tableofcontents
\renewcommand*{\thefootnote}{\arabic{footnote}}
\setcounter{footnote}{0}

\section{Introduction \label{sec: intro}}

The long-range Ising (LRI) model, with ferromagnetic couplings $J(r)\propto r^{-(d+s)}$, is the prototype of criticality with power-law interactions~\cite{Defenu:2021gzg}. Its interacting critical points are nonlocal conformal field theories (CFTs)~\cite{Paulos:2015jfa} in which the spin keeps its mean-field dimension $\Ds=(d-s)/2$, because the long-range coupling is not renormalized~\cite{Fisher:1972zz,Sak:1973oqx}. In two and three dimensions they cross over to the short-range Ising CFT for faster decay~\cite{Sak:1973oqx,Behan:2017dwr}. In one dimension they also describe quantum impurities: the localization transition of the sub-Ohmic spin-boson model, a two-level system coupled to a bosonic bath, lies in the one-dimensional LRI class~\cite{Leggett:1987zz,Winter:2008mmo,Shen:2023qtf}, the Ising-anisotropic Bose--Fermi Kondo model maps onto a chain with $s<1$~\cite{Kirchner:2008xx}, and the anisotropic Kondo problem onto the chain with $s=1$~\cite{Anderson:1970kyb}.

In every dimension the interacting critical points occupy a window of $s$. For $s<d/2$ the exponents are those of mean-field theory~\cite{Fisher:1972zz}. In $d=2,3$ the window ends at the crossover to the short-range Ising CFT, which Sak's criterion places at $s_*=2-\eta_{\rm SR}$~\cite{Sak:1973oqx,Behan:2017dwr}. In $d=1$ the model orders at finite temperature for $0<s<1$~\cite{Dyson:1968up}, and its window ends at $s=1$, where the inverse-square chain has a Kosterlitz--Thouless-like transition with a discontinuous magnetization~\cite{Thouless:1969zz,Frohlich:1982fs,Kosterlitz:1976zz,Aizenman:1987gn}. Each window thus holds a one-parameter family of interacting CFTs, and the one-dimensional family is probably the simplest such family not yet solved analytically~\cite{Benedetti:2024wgx}. Both ends are weakly coupled. Near the mean-field end the theory is a nonlocal $\phi^4$ theory, and the dimension $\DE$ of the leading $\mathbb Z_2$-even scalar $\mathcal E$ has an expansion in $\epsilon=2s-d$~\cite{Fisher:1972zz}. Near the short-range end, in $d=2,3$ the short-range Ising CFT couples weakly to a generalized free field~\cite{Behan:2017dwr,Behan:2017emf}, and in $d=1$ a recent strong--weak duality makes the end $s=1$ perturbative~\cite{Benedetti:2024wgx,Benedetti:2025nzp}. In between, neither description is weakly coupled, and computing the CFT data is a hard problem.

Existing methods capture only parts of this picture. The $\epsilon$ expansion was known to \PTorderLitword{} loops in general $d$~\cite{Benedetti:2020rrq}. About the short-range end, the leading short-range coefficient had been estimated in $d=3$ and shown to vanish in $d=2$~\cite{Behan:2017emf}, and the expansion in $d=1$ reached \SRorderLitword{} coefficients~\cite{Benedetti:2024wgx,Benedetti:2025nzp}. Monte Carlo simulations determined $\DE$ at a few couplings in one, two and three dimensions, mostly with percent-level errors~\cite{Luijten:1997th,Uzelac:2001lc,Tomita:2009mc,Picco:2012ak,Angelini:2014koy,Xiao:2025sk}. Functional renormalization-group studies lack a controlled error estimate~\cite{Defenu:2014bea,Pagni:2025ebi}. The conformal bootstrap has found kinks in two and three dimensions~\cite{Behan:2018hfx,Behan:2023ile} and, in $d=1$, bounds and some analytic data~\cite{Ghosh:2026gku}. Numerical renormalization-group (NRG) studies of the spin-boson model reach the window~\cite{Bulla:2003rns}, although truncation errors long spoiled their exponents in the mean-field regime $s<1/2$~\cite{Vojta:2012xx}.

In this article we study the LRI CFT in one, two and three dimensions from several independent directions. The $\epsilon$ expansion now reaches \PTorderOneDword{} loops in one, two and three dimensions, with several new coefficients in closed form. About the short-range end, the expansion in $d=1$ reaches \SRorderword{} coefficients, the first seven in closed form and the eighth in a conjectured closed form. In $d=2$ the leading nonvanishing short-range coefficient is now known in closed form, and in $d=3$ the estimate of the leading one is improved by computing it from the bootstrap spectrum of the short-range Ising CFT~\cite{Simmons-Duffin:2016wlq,Chang:2024whx}. We develop Hamiltonian truncation~\cite{Yurov:1989yu,Hogervorst:2014rta} for this nonlocal theory, whose nonlocality enters only through the generalized free field, a free theory with a Fock space on the cylinder. It reproduces low orders of the $\epsilon$ expansion without Feynman graphs and, once we correct a truncation artifact that persists at any cutoff, gives $\DE$ in $d=1$ at finite coupling. We also compute $\DE$ with NRG in $d=1$ and with cluster Monte Carlo (MC) simulations~\cite{Fukui:2009on} of the lattice model, and we resum the two expansions across the window in each dimension.

Universality provides a strong cross-check. A CFT does not depend on the microscopic model or the formalism used to reach it, so descriptions with unrelated sources of error must give the same scaling dimensions. Such a check matters even more in AI-driven research, whose long derivations and codes are hard to verify; agents can pass numerical checks while misidentifying a problem's structure~\cite{Zhao:2026sm}. The sharpest test is in $d=1$, where five descriptions apply, each simplest in one dimension or existing only there: the $\phi^4$ theory and its strong--weak dual, a Hamiltonian on the conformal cylinder, the spin-boson impurity and the lattice chain. Except near the mean-field end, where NRG does not converge, Hamiltonian truncation, NRG and MC now agree at the sub-percent level, and Hamiltonian truncation and NRG agree to within two parts in a thousand. The $\epsilon$ expansion uses the same graphs in every dimension, with coefficients analytic in $d$, so the one-dimensional checks also test the framework behind the two- and three-dimensional series.

\section{The long-range Ising model: one CFT, several descriptions \label{sec: model}}

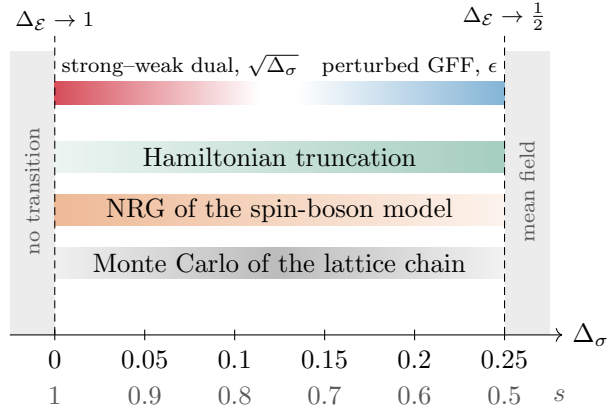
\begin{figure}[t]
\centering
\begin{tikzpicture}[x=0.85cm,y=1cm,font=\small]
 \fill[gray!15] (-0.7,0) rectangle (0,3.75);
 \fill[gray!15] (7,0) rectangle (7.7,3.75);
 \node[rotate=90,gray!70!black,font=\scriptsize] at (-0.35,1.9) {no transition};
 \node[rotate=90,gray!70!black,font=\scriptsize] at (7.35,1.9) {mean field};
 \shade[left color=cSR!80,right color=white] (0,3.05) rectangle (3.2,3.35);
 \shade[left color=white,right color=cEps!55] (3.8,3.05) rectangle (7,3.35);
 \node[anchor=south west,inner sep=1pt,font=\scriptsize] at (0.06,3.37) {strong--weak dual, $\sqrt{\Ds}$};
 \node[anchor=south east,inner sep=1pt,font=\scriptsize] at (6.94,3.37) {perturbed GFF, $\epsilon$};
 \shade[left color=cHT!8,right color=cHT!40] (0,2.15) rectangle (7,2.55);
 \node at (3.5,2.35) {Hamiltonian truncation};
 \shade[left color=cNRG!45,right color=cNRG!8] (0,1.45) rectangle (7,1.85);
 \node at (3.5,1.65) {NRG of the spin-boson model};
 \shade[left color=cMC!6,right color=cMC!6,middle color=cMC!30] (0,0.75) rectangle (7,1.15);
 \node at (3.5,0.95) {Monte Carlo of the lattice chain};
 \draw[densely dashed] (0,0) -- (0,3.95);
 \draw[densely dashed] (7,0) -- (7,3.95);
 \node[anchor=south,inner sep=2pt,font=\scriptsize] at (0,3.95) {$\DE\to1$};
 \node[anchor=south,inner sep=2pt,font=\scriptsize] at (7,3.95) {$\DE\to\tfrac12$};
 \draw[->] (-0.7,0) -- (7.9,0) node[right] {$\Ds$};
 \foreach \x/\lab in {0/0,1.4/0.05,2.8/0.1,4.2/0.15,5.6/0.2,7/0.25}
  \draw (\x,0.07) -- (\x,-0.07) node[below] {$\lab$};
 \foreach \x/\lab in {0/1,1.4/0.9,2.8/0.8,4.2/0.7,5.6/0.6,7/0.5}
  \node[gray!70!black] at (\x,-0.8) {$\lab$};
 \node[gray!70!black,right] at (7.6,-0.8) {$s$};
\end{tikzpicture}
\caption{Descriptions of the critical point in $d=1$, $J(r)\propto r^{-(1+s)}$, across the interacting window $0<\Ds<1/4$, or $1/2<s<1$ (lower scale). The $\epsilon$ expansion of the perturbed GFF is controlled at the mean-field end, and the expansion of the strong--weak dual in $\sqrt{\Ds}$ at the short-range end. Hamiltonian truncation, NRG and MC are nonperturbative and reach across the window. Shading fades, schematically, where each description loses accuracy. All describe one CFT, so they must give the same $\DE$. One dimension is the clearest test of universality, since all five descriptions apply there. In $d=2,3$ the short-range end is described by the expansion about the short-range Ising CFT instead of the dual, there is no NRG, and Hamiltonian truncation is extrapolated in its cutoff only order by order in $\epsilon$ (Sec.~\ref{sec: pert-gff}).}
\label{fig:formalisms}
\end{figure}

A critical point belongs to a universality class, not to a microscopic model. The LRI fixed point is a nonlocal CFT~\cite{Paulos:2015jfa}, and every model in its class shares its scaling dimensions. These depend only on $d$ and $s$, and the spin dimension $\Ds=(d-s)/2$ is exact. We obtain $\DE$, equivalently the RG eigenvalue $y_t=d-\DE$ of the coupling to $\mathcal E$ and the correlation-length exponent $\nu=1/y_t$, in the interacting window between the mean-field end $s=d/2$ and the short-range end,
\begin{equation}
 d=1:\ \tfrac12<s<1,\qquad d=2:\ 1<s<\tfrac74,\qquad d=3:\ \tfrac32<s<s_*\simeq1.9637,
 \label{eq:windows}
\end{equation}
that is $0<\Ds<\frac14$, $\frac18<\Ds<\frac12$ and $0.518<\Ds<\frac34$. In $d=2,3$ the upper end is $s_*=2-\eta_{\rm SR}$, by Sak's criterion~\cite{Sak:1973oqx,Behan:2017dwr}, with $\eta_{\rm SR}=\frac14$ in $d=2$ and from the bootstrap in $d=3$~\cite{Chang:2024whx}; in $d=1$ it is the inverse-square chain. We use five descriptions whose errors have different sources: two expansions, one from each end, two Hamiltonian methods and MC. The $\epsilon$ expansion, Hamiltonian truncation~\cite{Hogervorst:2014rta} and MC apply in every $d$. Near the short-range end the expansion takes a different form in $d=1$ and in $d=2,3$, and NRG exists only in $d=1$. Figure~\ref{fig:formalisms} shows all five in $d=1$: the two expansions are controlled at opposite ends, and neither reaches the other end without resummation.

\subsection{Lattice model and Monte Carlo \label{sec: model-mc}}

The microscopic model is a periodic hypercubic lattice of $N=L^d$ Ising spins $S_i=\pm1$,
\begin{equation}
 H=-\sum_{i<j}J_L(\mathbf r_i-\mathbf r_j)\,S_iS_j,\qquad J_L(\mathbf r)=\sum_{\mathbf m\in\mathbb Z^d}|\mathbf r+L\mathbf m|^{-(d+s)},
 \label{eq:chain}
\end{equation}
which keeps every periodic image. The Fukui--Todo algorithm~\cite{Fukui:2009on} performs Swendsen--Wang cluster updates~\cite{Swendsen:1987ce} at a cost per sweep linear in $N$. Because $\Ds$ is exact, $Y=L^{2\Ds}\langle m^2\rangle$, with $m$ the magnetization per spin, is scale invariant at criticality, like the Binder cumulant~\cite{Binder:1981sa}. Their temperature derivatives therefore grow as $L^{y_t}$.

\subsection{Perturbed generalized free field \label{sec: model-gff}}

For $s<2$ the continuum limit of $J(r)$ is a nonlocal kinetic term~\cite{Fisher:1972zz},
\begin{equation}
 S=\int d^dx\left(\frac{\mathcal N_s}{2}\,\phi\,(-\partial^2)^{s/2}\phi+\frac{r_0}{2}\phi^2+\frac{g_0}{4!}\phi^4\right),
 \label{eq:action}
\end{equation}
with $(-\partial^2)^{s/2}$ a nonlocal operator whose kernel falls as $|x-y|^{-(d+s)}$, and $\mathcal N_s$ fixed by $\langle\phi(x)\phi(0)\rangle_0=|x|^{-2\Ds}$. The free theory is a generalized free field (GFF), and $\phi^4$ has RG eigenvalue $d-4\Ds=\epsilon=2s-d$. The kinetic term receives no counterterm~\cite{Sak:1973oqx,Paulos:2015jfa,Benedetti:2020rrq}, so $\phi$ represents $\sigma$ with exact dimension $\Ds$. The beta function $\beta(g)=-\epsilon g+O(g^2)$ has an infrared fixed point at $g_*=O(\epsilon)$, where the anomalous dimension $\gamma_r$ of the mass term gives $\DE=2\Ds+\gamma_r(g_*)=\sum_nC_n(d)\,\epsilon^n$. The series is asymptotic, so it is controlled only near $s=d/2$.

\subsection{Short-range end \label{sec: model-sr}}

\textit{One dimension: strong--weak dual.} Near $s=1$ the domain walls of the spin chain form a dilute gas of alternating kinks and antikinks~\cite{Anderson:1970kyb,Kosterlitz:1976zz}. Benedetti et al.\ recast this gas as a qubit coupled to a compact GFF $\theta\sim\theta+2\pi/b_0$ of negative dimension $-\Ds$, with $\langle[\theta(x)-\theta(0)]^2\rangle_0\propto|x|^{2\Ds}$~\cite{Benedetti:2024wgx,Benedetti:2025nzp}:
\begin{equation}
 Z=\Big\langle\mathrm{tr}\,\mathrm{P}\exp\int dx\Big[g\left(\hat\sigma_+e^{\ii b_0\theta}+\hat\sigma_-e^{-\ii b_0\theta}\right)+\frac{\ii h}{\sqrt2}\,\hat\sigma_z\,\partial_x\theta\Big]\Big\rangle_0,
 \label{eq:dual}
\end{equation}
with $\hat\sigma_\pm=(\hat\sigma_x\pm\ii\hat\sigma_y)/2$. Since $\hat\sigma_\pm^2=0$, the expansion in the fugacity $g$ reproduces the alternating gas. The one-dimensional LRI CFT is the sector of the infrared fixed point invariant under a $U(1)$ that shifts $\theta$ and rotates the qubit, with the spin field $\sigma$ represented by $\hat\sigma_z$. There $g_*=O(\Ds^{1/2})$, so the dual is weakly coupled where $\phi^4$ is not. In the flow of $g$ and $b^2$, with $h$ the coupling of $\hat\sigma_z\partial_x\theta$ and $b=b_0-\sqrt2h$ the effective charge, one combination of the two perturbations, $\mathcal O_-$, is relevant. Assuming no level crossing, $\mathcal O_-$ is identified with $\mathcal E$, so $\DE=1+\sum_{n\geq1}C^{\rm SR}_n(1)\Ds^{n/2}$ with $y_t\simeq2\sqrt{\Ds}$~\cite{Kosterlitz:1976zz,Benedetti:2024wgx}.

\textit{Two and three dimensions: short-range Ising CFT coupled to a GFF.} Near $s_*$ the fixed point is the short-range Ising (SRI) CFT coupled to a GFF $\chi$ of dimension $(d+s)/2$~\cite{Behan:2017dwr,Behan:2017emf},
\begin{equation}
 S=S_{\rm SRI}+S_\chi+g\int d^dx\,\sigma\chi,
 \label{eq:SRIchi}
\end{equation}
with $\sigma$ the spin field of the SRI CFT and $S_\chi$ the Gaussian action of $\chi$, normalized by $\langle\chi(x)\chi(0)\rangle=|x|^{-(d+s)}$. The coupling has dimension $\delta=(s_*-s)/2$, its fixed point is at $g_*^2=O(\delta)$, and $\DE=\DE^{\rm SR}+\sum_{n\geq1}C^{\rm SR}_n(d)\,\delta^n$, with coefficients given by integrated correlators of the SRI CFT. This description shares no ingredient with the $\epsilon$ expansion.

\subsection{Hamiltonian truncation \label{sec: model-ht}}

Radial quantization maps the GFF to the cylinder $\mathbb R\times S^{d-1}$, where it becomes a set of free oscillators $a_k$ with energies $\Ds+\kappa_k$, the integers $\kappa_k$ labeling descendants. In $d=1$, where the sphere is two points, there is one mode for each $\kappa=n\geq0$; in $d=2$ the modes are labeled by $a,b\geq0$, with $\kappa=a+b$ and spin $a-b$ around the circle; in $d=3$ they carry the spherical harmonics $Y_{lm}$, with $\kappa=2n+l$ and $n\geq0$. Each quantum carries energy $\Ds$ plus the $\kappa_k$ of its mode, so a Fock state with $P$ quanta has free energy $P\Ds+\ell$, where its level $\ell$ is the sum of the $\kappa_k$ alone. As in the truncated conformal space approach~\cite{Yurov:1989yu,Hogervorst:2014rta} and related conformal truncations~\cite{Anand:2020gnn}, we add the interactions of Eq.~\eqref{eq:action} on the sphere and diagonalize
\begin{equation}
 H=\sum_k(\Ds+\kappa_k)\,a_k^\dagger a_k+\Big\langle r\,{:}\phi^2{:}+\frac{g}{12}\,{:}\phi^4{:}\Big\rangle_{S^{d-1}}
 \label{eq:HT}
\end{equation}
in a truncated Fock space, with $\langle\cdot\rangle$ the average over the sphere. Its eigenvalues, measured from the vacuum energy, estimate scaling dimensions. At each cutoff, $r$ and $g$ are tuned to two exact properties of the CFT: $\phi$ has dimension $\Ds$, and $\phi^3$ has the shadow dimension $d-\Ds$ by the nonlocal equation of motion~\cite{Paulos:2015jfa,Benedetti:2024wgx}. The first excited $\mathbb Z_2$-even gap, $\DE$, is then a prediction, and the descendant spacings test conformality. Expanded in $\epsilon$, the method reproduces the low-order $C_n(d)$ without Feynman graphs; at finite coupling, with the levels cut at $\ell\leq N$, it gives $\DE$ in $d=1$. Both rest on one key improvement. A cutoff leaves an excited state fewer vacuum bubbles than the vacuum, because the state's own quanta use up part of the cutoff~\cite{Rychkov:2014eea}, and here the difference does not vanish as the cutoff grows. We correct this truncation artifact by adding to the Hamiltonian a term built from the vacuum energies at lower cutoffs, with no fitted parameter, a nonperturbative analogue of the state-dependent counterterms of Ref.~\cite{Elias-Miro:2020qwz}. Appendix~\ref{app: window} explains the artifact and its correction in $d=1$, including the treatment of the zero mode, and the \SM{} (Parts~III and~IV) gives the construction in every dimension.

\subsection{Spin-boson model and NRG \label{sec: model-nrg}}

In $d=1$ a quantum impurity gives a further description. The sub-Ohmic spin-boson model couples a two-level system with tunneling $\Omega$ to a bosonic bath,
\begin{equation}
 H_{\rm SB}=-\frac{\Omega}{2}\hat\sigma_x+\sum_i\omega_i\,c_i^\dagger c_i+\frac{\hat\sigma_z}{2}\sum_i\lambda_i\left(c_i+c_i^\dagger\right),
 \label{eq:SB}
\end{equation}
with spectral density $\pi\sum_i\lambda_i^2\,\delta(\omega-\omega_i)\propto\omega^{s}$. Integrating out the bath leaves an imaginary-time Ising chain $\hat\sigma_z(\tau)$ with interaction $\propto\tau^{-(1+s)}$ and the same $s$, whose inverse temperature is the dissipation strength~\cite{Leggett:1987zz,Winter:2008mmo}. The localization transition is thus in the $d=1$ LRI class, with $\hat\sigma_z$ representing $\sigma$; the mapping has no analogue for $d>1$. The bosonic NRG~\cite{Bulla:2003rns} maps the logarithmically discretized bath onto a Wilson chain, diagonalized iteratively in a truncated basis. Each site lowers the energy scale by a fixed factor $\Lambda$, so the iteration is an RG flow. At criticality, the response of the rescaled levels to the dissipation strength grows as $\Lambda^{y_tn}$ at site $n$. Reaching exponentially small scales, NRG remains usable close to $s=1$, where $y_t\to0$, but loses accuracy toward $s=1/2$.

\section{Results \label{sec: pert}}

Most of these results rest on the ability of frontier AI models to carry out long perturbative calculations correctly, both for the coefficients themselves and for the exact asymptotic forms that guide the numerical methods: the large-cutoff behavior used to extrapolate Hamiltonian truncation and the large-order behavior used to resum the series. The \SM{} opens with an account of how the work was organized and checked. We first report the two expansions, about the generalized free field (Sec.~\ref{sec: pert-gff}) and about the short-range end (Sec.~\ref{sec: sr}), and resum them into one curve across each window (Sec.~\ref{sec: resum}). We then report the nonperturbative values (Sec.~\ref{sec: nonpert}) and compare all descriptions (Sec.~\ref{sec: compare}). Appendix~\ref{app: closed} gives the closed forms and Appendix~\ref{app: numerics} the numerical values; the detailed methods are given in the \SM{}, the file \texttt{Supplementary Material.pdf} of the release at \url{https://github.com/Canonical111/LRI_public}, which also holds the code that reproduces every result (see Data and code availability). Throughout this article, uncertainties are rounded by the rule of the Particle Data Group~\cite{ParticleDataGroup:2024cfk}, with two exceptions: values from other work are quoted as published, and our MC fit results are given to five decimal places, with both their errors in units of $10^{-5}$.

\subsection{Around the generalized free field \label{sec: pert-gff}}

About the generalized free field, $\DE$ is a power series in $\epsilon=2s-d$,
\begin{equation}
 \DE=\sum_{n\geq0}C_n(d)\,\epsilon^n ,
 \label{eq:eps-series}
\end{equation}
and Table~\ref{tab:pt} gives the coefficients $C_n(d)$. The published series reaches three loops in general $d$~\cite{Benedetti:2020rrq}; we extend it to \PTorderOneDword{} loops in $d=1,2,3$. Through five loops each coefficient is an exact expression, valid for general $d$, in a few master integrals, and the masters are evaluated separately in each dimension (Appendix~\ref{app: closed} and \SM{}, Part~I, Sec.~1.1). In $d=1$, $C_4$ and $C_5$ are finite combinations of Gamma and polygamma values and of seven ordinary hypergeometric values, and in $d=2$ the same values, with Catalan's constant and $\zeta(3)$, give $C_5(2)$ in closed form. In $d=3$, $C_3$, $C_4$ and $C_5$ contain values of hypergeometric functions of two variables, $C_3$ through the tetrahedron graph; we evaluate these values to \PTdThreeDigits{} digits but do not reduce them to functions of one variable. From six loops on the coefficients are numerical, apart from a representation of $C_6(1)$ in nine special-function values. They alternate in sign and grow factorially, as expected for an asymptotic series: in $d=1$ the magnitude of $C_{n+1}/C_n$ grows steadily, from 12 at $n=3$ to 25 at $n=7$.

\begin{table}[h]
\centering
\renewcommand{\arraystretch}{1.25}
\begin{tabular}{@{}c l l l@{}}
\toprule
Order & \multicolumn{1}{c}{$d=1$} & \multicolumn{1}{c}{$d=2$} & \multicolumn{1}{c}{$d=3$}\\
\midrule
$C_0$ & \ptPub{1/2} & \ptPub{1} & \ptPub{3/2}\\
$C_1$ & \ptPub{-1/6} & \ptPub{-1/6} & \ptPub{-1/6}\\
$C_2$ & \ptPub{\CtwoOne} & \ptPub{\CtwoTwo} & \ptPub{\CtwoThree}\\
$C_3$ & \ptPub{\CthreeOne} & \ptPub{\CthreeTwo} & \ptPub{\CthreeThree}\\
$C_4$ & \ptClosed{\CfourOne} & \ptClosed{\CfourTwo} & \ptExact{\CfourThree}\\
$C_5$ & \ptClosed{\CfiveOne} & \ptClosed{\CfiveTwo} & \ptExact{\CfiveThree}\\
$C_6$ & \ptExact{\CsixOne} & \ptNum{\CsixTwo} & \ptNum{\CsixThree}\\
$C_7$ & \ptNum{\CsevenOne} & \ptNum{\CsevenTwo} & \ptNum{\CsevenThree}\\
$C_8$ & \ptNum{\CeightOne} & \ptNum{\CeightTwo} & \ptNum{\CeightThree}\\
\bottomrule
\end{tabular}
\caption{Coefficients $C_n(d)$ of Eq.~\eqref{eq:eps-series}. Black: published, through three loops~\cite{Benedetti:2020rrq}. \textcolor{ptnew}{Blue, $\ast$}: new, in closed form. \textcolor{ptnew}{Blue, $\ddagger$}: new, in special-function values that are not reduced to standard constants (Sec.~\ref{sec: pert-gff}). \textcolor{ptnum}{Red, $\dagger$}: new, numerical. Analytic entries are rounded. Numerical uncertainties are mostly four sampling standard errors.}
\label{tab:pt}
\end{table}

\textit{Independent checks.} The numerical uncertainties do not cover a mistake shared by the whole graph and renormalization framework; only the other descriptions can test that. Since $C_n(d)$ is analytic in $d$, with graphs and renormalization structure common to all dimensions, the one-dimensional comparisons (Sec.~\ref{sec: compare}) also test the framework behind $C_n(2)$ and $C_n(3)$, though not their $d$-specific integrals and subtractions. Hamiltonian truncation tests the low orders without Feynman graphs (Table~\ref{tab:htpert}): expanded in $\epsilon$, the truncated Hamiltonian~\eqref{eq:HT} gives every coefficient at finite cutoff, and, extrapolated in the cutoff with the truncation artifact corrected (Sec.~\ref{sec: model-ht}), it reproduces the coefficients of Table~\ref{tab:pt}. In $d=1$, with a cutoff $\Lambda$ on the energy, exact sums over intermediate states to $\Lambda=\HTpOneLexact$ ($\NOneCfiveLmax$ for $C_5$) give $C_2$ to $C_5$ to $\HTpCfiveOneDev$ or better, and the truncated spectrum to $\Lambda=\HTpLamThermal$ ($\HTpLamCeight$ for $C_8$) gives $C_6$ to $C_8$ to $\HTpCsevenOneDev$. In $d=2$, with a cutoff $L$ on the level, $C_2$ to $C_5$ agree to $\HTpCfiveTwoDev$ or better, $C_2$ to $C_4$ to better than $\HTpTwoLowDev$. In $d=3$, with the same level cutoff, exact sums to $L=\HTpThreeLfour$ give $C_2$ to $C_4$ to $\HTpCfourThreeDev$ or better.

\begin{table}[tbp]
\centering
\renewcommand{\arraystretch}{1.25}
\small
\setlength{\tabcolsep}{4pt}
\begin{tabular}{@{}c l l l l l l@{}}
\toprule
 & \multicolumn{2}{c}{$d=1$} & \multicolumn{2}{c}{$d=2$} & \multicolumn{2}{c}{$d=3$}\\
\cmidrule(lr){2-3}\cmidrule(lr){4-5}\cmidrule(l){6-7}
Order & \multicolumn{1}{c}{HT} & \multicolumn{1}{c}{$\delta$} & \multicolumn{1}{c}{HT} & \multicolumn{1}{c}{$\delta$} & \multicolumn{1}{c}{HT} & \multicolumn{1}{c}{$\delta$}\\
\midrule
$C_2$ & \ptPub{\HTpCtwoOne} & $\HTpCtwoOneDev$ & \ptPub{\HTpCtwoTwo} & \HTpCtwoTwoDigits{} digits & \ptPub{\HTpCtwoThree} & \HTpCtwoThreeDigits{} digits\\
$C_3$ & \ptPub{\HTpCthreeOne} & $\HTpCthreeOneDev$ & \ptPub{\HTpCthreeTwo} & $\HTpCthreeTwoDev$ & \ptPub{\HTpCthreeThree} & $\HTpCthreeThreeDev$\\
$C_4$ & \ptClosed{\HTpCfourOne} & $\HTpCfourOneDev$ & \ptClosed{\HTpCfourTwo} & $\HTpCfourTwoDev$ & \ptExact{\HTpCfourThree} & $\HTpCfourThreeDev$\\
$C_5$ & \ptClosed{\HTpCfiveOne} & $\HTpCfiveOneDev$ & \ptClosed{\HTpCfiveTwo} & $\HTpCfiveTwoDev$ & \ptExact{\HTpCfiveThree} & $\HTpCfiveThreeDev$\\
$C_6$ & \ptExact{\HTpCsixOne} & $\HTpCsixOneDev$ & n.r. & & n.r. & \\
$C_7$ & \ptNum{\HTpCsevenOne} & $\HTpCsevenOneDev$ & n.r. & & n.r. & \\
$C_8$ & \ptNum{\HTpCeightOne} & $\HTpCeightOneDev$ & n.r. & & n.r. & \\
\midrule
$C^{\rm SR}_2(2)$ & & & \ptClosed{\SRtwoCtwoHT} & $\SRtwoCtwoHTDev$ & & \\
\bottomrule
\end{tabular}
\caption{Perturbative Hamiltonian truncation (HT): $C_n(d)$ and $\delta=|C_n^{\rm HT}/C_n-1|$ against Table~\ref{tab:pt}, and the short-range coefficient $C^{\rm SR}_2(2)$ against Table~\ref{tab:sr}. Colors and markers are those of the reference values in Tables~\ref{tab:pt} and~\ref{tab:sr}; for a range, $\delta$ is its largest deviation. The ranges and errors measure the spread over extrapolation fits, except for $C^{\rm SR}_2(2)$, whose error sums systematic diagnostics; none is a standard deviation. In $d=3$, $C_5$ is provisional, from sums to $L=\HTpThreeLfive$. In $d=1$, $C_8$ is calibrated on $C_6$ and $C_7$; without that calibration the extrapolations give $(\HTpCeightOneUncal)\times10^6$, which contains the value of Table~\ref{tab:pt}. Coefficients marked n.r.\ are not reproduced: at these orders the cutoffs within reach of the available memory are too small to extrapolate.}
\label{tab:htpert}
\end{table}

\textit{Higher orders.} The same methods can reach the ninth order, but it would sharpen the resummed curves of Sec.~\ref{sec: resum} only if it were very precise. Re-expanding the eighth-order resummations in $\epsilon$ already gives $C_9(1)$ between $\RSnineImplied$, so a computed value adds information only if its error is below this spread of about $\RSnineSpread\%$. At $\Ds=0.1$, a change of 1\% in $C_9(1)$ moves the curve by $\RSnineOnePct$, and adding $C_9$ makes the curve about \RSnineCeightAmp{} times more sensitive to the error of $C_8(1)$, so $C_8$ would have to be improved as well. Each order also multiplies the number of graphs and the cost of sampling them; the eighth order in $d=1$ alone took about $10^5$ core-hours. A useful ninth order is therefore possible but very costly.

\subsection{Near the short-range end \label{sec: sr}}

Near the short-range end, $\DE$ is a power series in $\sqrt{\Ds}$ in $d=1$ and in $\delta=(s_*-s)/2$ in $d=2,3$ (Sec.~\ref{sec: model-sr}),
\begin{equation}
 d=1:\ \DE=1+\sum_{n\geq1}C^{\rm SR}_n(1)\,\Ds^{n/2},\qquad d=2,3:\ \DE=\DE^{\rm SR}+\sum_{n\geq1}C^{\rm SR}_n(d)\,\delta^n ,
 \label{eq:sr-series}
\end{equation}
and Table~\ref{tab:sr} gives the coefficients.

\textit{One dimension.} The published expansion of the dual has two coefficients, $C^{\rm SR}_1(1)=-2$ and $C^{\rm SR}_2(1)=\frac12$~\cite{Benedetti:2024wgx,Benedetti:2025nzp}. We computed $C^{\rm SR}_3(1)$ to $C^{\rm SR}_8(1)$. Through $C^{\rm SR}_7(1)$ the coefficients are proved to be rational combinations of zeta values,
\begin{equation}
\begin{aligned}
 C^{\rm SR}_3(1)&=\frac{39}{16}+2\zeta(2),\qquad C^{\rm SR}_4(1)=4-4\zeta(2)-8\zeta(3),\\
 C^{\rm SR}_5(1)&=\frac{11249}{1024}-\frac{361}{16}\zeta(2)+24\zeta(3)+\frac{49}{2}\zeta(4),\\
 C^{\rm SR}_6(1)&=26-38\zeta(2)+104\zeta(3)-80\zeta(4)-272\zeta(5)+112\zeta(2)\zeta(3),\\
 C^{\rm SR}_7(1)&=\frac{1798583}{32768}-\frac{90637}{1024}\zeta(2)+\frac{453}{4}\zeta(3)-\frac{16835}{64}\zeta(4)+1224\zeta(5)\\
 &\quad-540\zeta(2)\zeta(3)+\frac{4335}{8}\zeta(6)-320\zeta(3)^2.
\end{aligned}
\label{eq:sr-closed}
\end{equation}
For $C^{\rm SR}_8(1)$ we find the closed form
\begin{equation}
\begin{aligned}
 C^{\rm SR}_8(1)={}&88-178\zeta(2)+110\zeta(3)+113\zeta(4)+4892\zeta(5)-2376\zeta(2)\zeta(3)-\frac{8630}{3}\zeta(6)+1872\zeta(3)^2\\
 &-12168\zeta(7)+6048\zeta(2)\zeta(5)+1536\zeta(3)\zeta(4)
\end{aligned}
\label{eq:sr-C8}
\end{equation}
It is conjectured: it agrees with the numerical value of $C^{\rm SR}_8(1)$ to about 85 digits and reproduces its rational term, 88, which is computed exactly.

\textit{Two and three dimensions.} In $d=2$ the first coefficient vanishes, $C^{\rm SR}_1(2)=0$~\cite{Behan:2017emf}. We derived $C^{\rm SR}_2(2)$ in closed form from the exact correlators of the two-dimensional Ising CFT~\cite{Ardonne:2010hj}, in Gamma values and three ${}_4F_3(1)$ constants (Appendix~\ref{cf:sec:cr2}). In $d=3$ we computed $C^{\rm SR}_1(3)=\SRthreeCone$, with an estimated range $\SRthreeConeRange$ that holds if the spectrum of Ref.~\cite{Simmons-Duffin:2016wlq} misses no operator below assumed gaps, compared with the estimate $\SRthreeConeBRRZ$ of Ref.~\cite{Behan:2017emf}.

\textit{Independent checks.} The new coefficients are also obtained by independent calculations. In $d=1$, every ingredient of $C^{\rm SR}_5(1)$ to $C^{\rm SR}_7(1)$ was recomputed from scratch, with matching results. A separate calculation starts from the Bethe-ansatz solution of the anisotropic Kondo model~\cite{Andrei:1982cr}. It reproduces the closed forms of $C^{\rm SR}_3(1)$ to $C^{\rm SR}_7(1)$ to a relative $\SRintegClosedDev$ and gives $C^{\rm SR}_8(1)=\SRCeightMag$, $\SRCeightMagDiff$ from the closed form~\eqref{eq:sr-C8}. In $d=2$, a Monte Carlo integration gives $C^{\rm SR}_2(2)=\SRtwoCtwoMC$, and fourth-order perturbation theory with the truncated cylinder Hamiltonian gives $\SRtwoCtwoHT$.

\begin{table}[h]
\centering
\renewcommand{\arraystretch}{1.25}
\setlength{\tabcolsep}{4pt}
\begin{tabular}{@{}l l l l l l l l l@{}}
\toprule
 & $n=1$ & 2 & 3 & 4 & 5 & 6 & 7 & 8\\
\midrule
$C^{\rm SR}_n(1)$ & \ptPub{-2} & \ptPub{1/2} & \ptClosed{\SRCthree} & \ptClosed{\SRCfour} & \ptClosed{\SRCfive} & \ptClosed{\SRCsix} & \ptClosed{\SRCseven} & \ptConj{\SRCeight}\\
$C^{\rm SR}_n(2)$ & \ptPub{0} & \ptClosed{\SRtwoCtwo} & & & & & & \\
$C^{\rm SR}_n(3)$ & \ptNum{\SRthreeCone} & & & & & & & \\
\bottomrule
\end{tabular}
\caption{Coefficients $C^{\rm SR}_n(d)$ of Eq.~\eqref{eq:sr-series} near the short-range end. Black: published~\cite{Benedetti:2024wgx,Benedetti:2025nzp,Behan:2017emf}. \textcolor{ptnew}{Blue, $\ast$}: new, in closed form (zeta values in $d=1$, Eq.~\eqref{eq:sr-closed}; Gamma values and ${}_4F_3(1)$ constants in $d=2$). \textcolor{ptnew}{Blue, \S}: new, in a conjectured closed form, Eq.~\eqref{eq:sr-C8}, which agrees with the numerical evaluation to about 85 digits. \textcolor{ptnum}{Red, $\dagger$}: numerical values. $C^{\rm SR}_1(3)$ improves the estimate $\SRthreeConeBRRZ$ of Ref.~\cite{Behan:2017emf}: it is computed from the bootstrap spectrum of the short-range Ising CFT~\cite{Simmons-Duffin:2016wlq,Chang:2024whx}, with the estimated range $\SRthreeConeRange$ valid if that spectrum misses no operator below assumed gaps.}
\label{tab:sr}
\end{table}

\subsection{Resummation \label{sec: resum}}

The $\epsilon$ series is asymptotic, and the window reaches $\epsilon=1$, $\frac32$ and $0.93$ at the short-range end in $d=1,2,3$. We resum it by conformal Borel resummation~\cite{LeGuillou:1979ixc}, which needs the location of the leading singularity of the Borel transform, and impose the short-range data of Sec.~\ref{sec: sr} at the far end of the window. This gives a curve $\DE(\Ds)$ across each window.

\begin{figure*}[t]
\centering
\includegraphics[width=\textwidth]{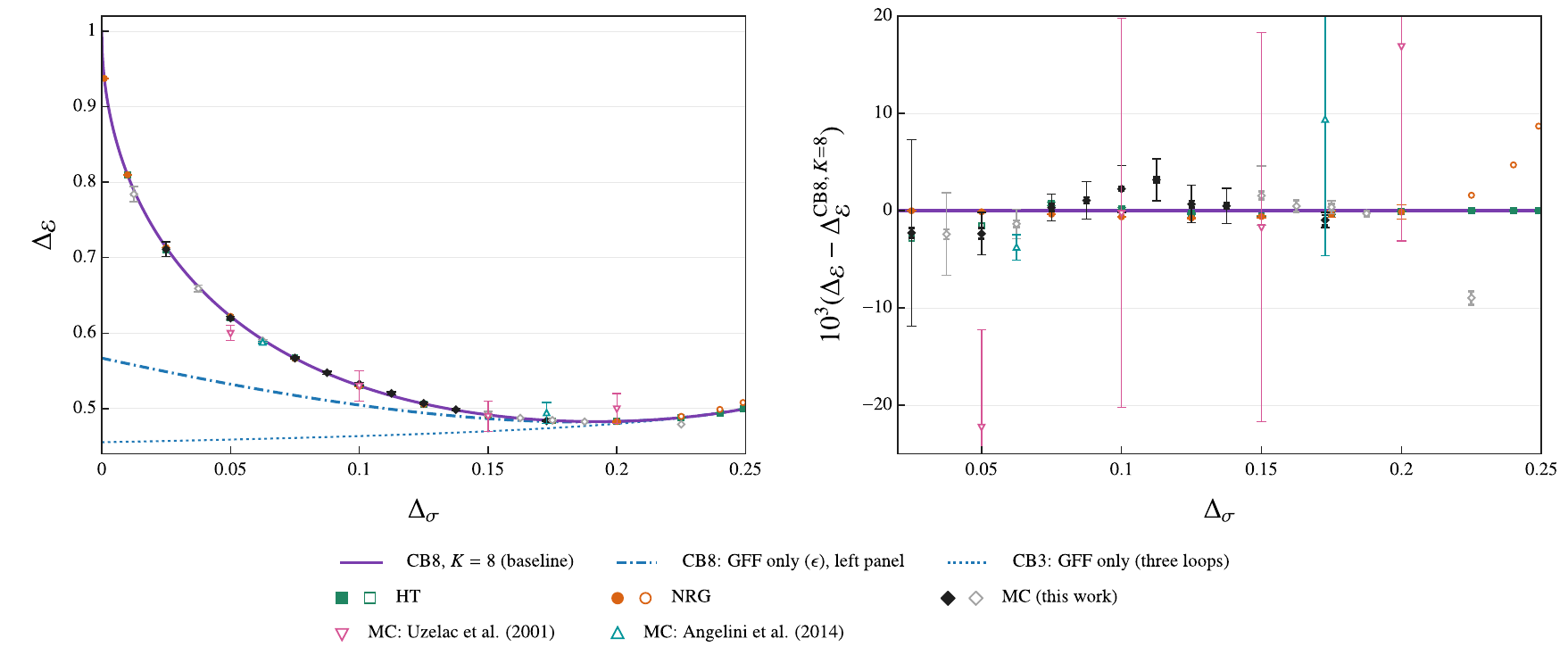}
\caption{$\DE$ in $d=1$. Left: $\DE$ across the window; right: the residual $10^3(\DE-\DE^{\rm res})$ from the resummed curve, for $0.02\le\Ds\le0.25$. Purple: the resummed curve (Sec.~\ref{sec: resum}), with eight short-range coefficients imposed (CB8, $K=8$ in the legend: eight loops, eight coefficients). Blue, left panel: the $\epsilon$ series alone, resummed at eight loops (dash-dot) and three loops (dotted). Green squares: Hamiltonian truncation, extrapolated in the cutoff (filled) or at its largest cutoff where it cannot be extrapolated (open). Orange circles: NRG, converged in chain length (filled) or not (open). Diamonds: MC of this work, accepted (black) or not (gray, open; Sec.~\ref{sec: nonpert}), with the statistical error (inner) and the fit-model half-range (outer). Open triangles: published MC~\cite{Uzelac:2001lc,Angelini:2014koy}, one color per reference (legend), with quoted errors; three of their bars are cut off in the right panel. Except the MC statistical errors, no bar or range is a standard deviation.}
\label{fig:cmp1d}
\end{figure*}

\begin{figure*}[t]
\centering
\includegraphics[width=\textwidth]{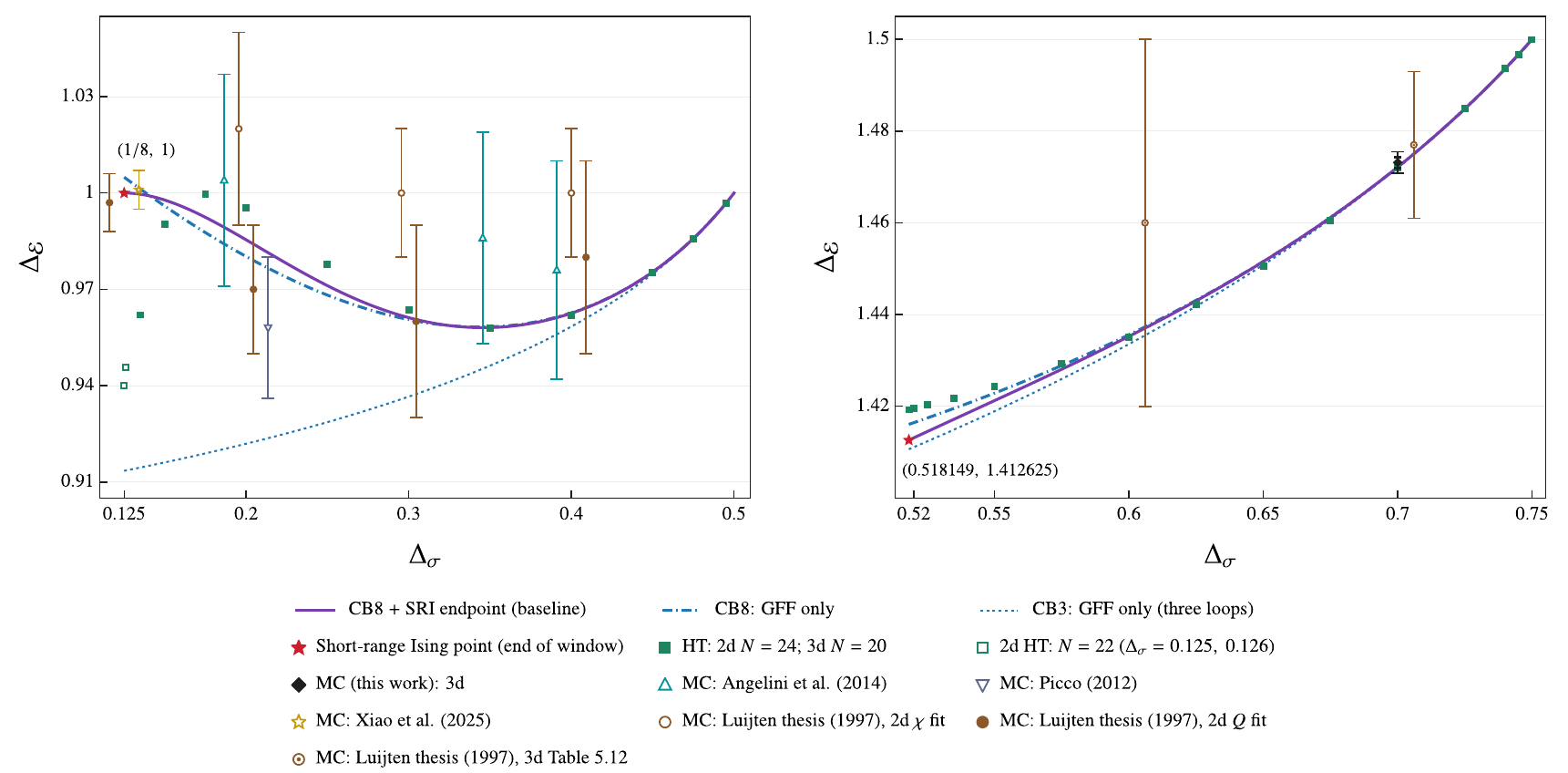}
\caption{$\DE$ in $d=2$ (left) and $d=3$ (right). Purple: the resummed curve of Sec.~\ref{sec: resum}, which imposes the short-range Ising end, marked by a red star (exact in $d=2$, from the bootstrap~\cite{Chang:2024whx} in $d=3$). Blue: the $\epsilon$ series alone, resummed at eight loops (dash-dot, CB8 in the legend) and three loops (dotted, CB3). Green squares: Hamiltonian truncation at the largest cutoff, not extrapolated; open squares mark $\Ds=0.125$ and 0.126, where the largest cutoff is smaller. Black diamond: the $d=3$ MC value, accepted under an amended fit protocol (Sec.~\ref{sec: nonpert}), with statistical (inner) and fit-model (outer) bars. Other symbols: published MC~\cite{Angelini:2014koy,Picco:2012ak,Xiao:2025sk,Luijten:1997th}, one color and shape per reference (legend), with quoted errors.}
\label{fig:cmp2d3d}
\end{figure*}

\textit{Large orders.} At negative coupling and $s=d/2$, the action~\eqref{eq:action} has a real instanton, the nonzero real stationary field of least action among finite-action fields. As in the large-order analysis of $\phi^4$ theory~\cite{Lipatov:1976ny,Brezin:1976vw}, it places the candidate singularity at $-A_\epsilon$, with
\begin{equation}
 A_\epsilon(1)=\frac{9\pi^2}{\Gamma(\frac14)^4},\qquad A_\epsilon(2)=\frac98,\qquad A_\epsilon(3)=\frac94\left[\frac{\Gamma(\frac94)}{\Gamma(\frac34)}\right]^2,
 \label{eq:Aeps}
\end{equation}
and power counting gives $C_n\propto(-1/A_\epsilon)^n\,\Gamma\big(n+\frac{d+7}2\big)$. In $d=1$ we compute the one-loop amplitude,
\begin{equation}
 C_n(1)\simeq K_I\left(-\frac1{A_\epsilon}\right)^n\Gamma(n+4),\qquad K_I=\RSKI .
 \label{eq:KI}
\end{equation}

\textit{One dimension.} Near $s=1$, $\DE-1\simeq-2\sqrt{\Ds}$ has a square-root branch point, which no finite-order Borel approximant in $\epsilon$ can produce: the resummed series alone ends at \RSgffOnlyEnd{} instead of 1. We therefore resum in $v=1-2\sqrt{\Ds}$, which removes the branch point, with the short-range series through $C^{\rm SR}_8(1)$ imposed. The curve falls from 1 at $\Ds=0$ to a shallow minimum \RSoneDmin{} at $\Ds=\RSoneDminAt$ and rises to $\frac12$ (Fig.~\ref{fig:cmp1d}).

\textit{Two and three dimensions.} Here $\DE$ is a power series in $s_*-s$ near the short-range end, Eq.~\eqref{eq:sr-series}, with no branch point. We therefore resum in $\epsilon$ directly, with the end-point value and the leading short-range coefficients imposed: in $d=2$ the vanishing slope and the curvature $C^{\rm SR}_2(2)$, in $d=3$ the slope $C^{\rm SR}_1(3)$ (Fig.~\ref{fig:cmp2d3d}). Without short-range input the eighth-order series already ends at \RSgffOnlyTwo{} and \RSgffOnlyThree, against 1 and the bootstrap value 1.4126~\cite{Chang:2024whx}, consistent with Sak's criterion~\cite{Sak:1973oqx,Behan:2017dwr,Behan:2017emf}.

\subsection{Nonperturbative results \label{sec: nonpert}}

Figures~\ref{fig:cmp1d} and~\ref{fig:cmp2d3d} show the nonperturbative values with the resummed curves of Sec.~\ref{sec: resum}; Appendix~\ref{app: numerics} lists them.

\textit{Hamiltonian truncation.} In $d=1$, the truncated cylinder Hamiltonian at finite coupling, with the truncation artifact corrected (Sec.~\ref{sec: model-ht}), gives $\DE$ extrapolated from level cutoffs $N\leq\HTnpNmax$ (Table~\ref{tab:num1}). The extrapolation is final for $0.1\leq\Ds\leq0.249$. At $\Ds\leq0.025$ the corrected values do not converge in $N$, a property of the correction rather than of the size of the truncated space (Appendix~\ref{app: window}). Near $s=1/2$ the truncation reproduces the $\epsilon$ expansion, to $\HTnpEpsDiff$ at $\Ds=0.249$. In $d=2,3$ we reach $N=\HTnpDtwoNmax$ and $\HTnpDthreeNmax$, too few cutoffs to extrapolate; Fig.~\ref{fig:cmp2d3d} shows finite-cutoff values, which describe the truncated model (Tables~\ref{tab:num2} and~\ref{tab:num3}). Expanded in $\epsilon$, the same truncation reproduces low orders of Table~\ref{tab:pt} without Feynman graphs (Sec.~\ref{sec: pert-gff}).

\textit{NRG.} In $d=1$, NRG gives $\DE$ at \NRGnPoints{} couplings (Table~\ref{tab:num1}). The three points closest to $s=1/2$ are not converged in chain length and enter no comparison. Near $s=1$, NRG reproduces the dual expansion of Table~\ref{tab:sr} through $C^{\rm SR}_5(1)$; the term of $C^{\rm SR}_6(1)$ is of the size of the NRG error, and the higher ones are smaller. A blind second implementation reproduces the Wilson chain and the estimates.

\textit{Monte Carlo.} The cluster simulations of the lattice model~\eqref{eq:chain} cover 20 couplings in $d=1$ and eight each in $d=2,3$ (Tables~\ref{tab:num1} to~\ref{tab:num3}), on lattices of linear size $L=\MConeL$, $\MCtwoL$ and $\MCthreeL$ in $d=1$, 2 and 3. A fit protocol accepts or rejects each value (\SM{}, Part~VI). In $d=1$, \MCnAcceptedword{} values are accepted. In $d=2$ no infinite-size value is accepted, and the robust result is the effective exponent from the two largest lattices. In $d=3$ one value is accepted, $\DE=\MCthreeD$ at $s=1.6$, under an amended protocol.

\subsection{Comparison \label{sec: compare}}

\textit{One dimension.} Away from the mean-field end, Hamiltonian truncation, NRG and MC agree at the sub-percent level (Fig.~\ref{fig:cmp1d}). Hamiltonian truncation and NRG agree to within two parts in a thousand wherever the truncation is extrapolated and NRG is converged in chain length, $0.075\leq\Ds\leq0.2$.

The resummed curve, which uses none of these values, lies within $\CMPhtCB$ of every final value of Hamiltonian truncation, within $\CMPnrgCB$ of every converged NRG value and within $\CMPmcCB$ of every accepted MC value. At the two ends the comparison rests on the expansions. Near $\Ds=1/4$ the $\epsilon$ series is accurate, NRG is not converged in chain length, and the MC estimates from finite lattices lie $\CMPmcGauss$ below the series. Near $\Ds=0$, the corrected truncation does not converge as the cutoff grows, and NRG reproduces the dual expansion through $C^{\rm SR}_5(1)$ (Sec.~\ref{sec: nonpert}). The comparison cannot distinguish resummed curves that impose four or eight short-range coefficients, which differ by at most $\CMPkfourKeight$.

\textit{Two and three dimensions.} The checks are weaker (Fig.~\ref{fig:cmp2d3d}). In $d=2$ no nonperturbative value is extrapolated: Hamiltonian truncation reaches too few cutoffs, and no infinite-size MC value is accepted. The effective MC exponents from the two largest lattices agree with the resummed curve to $\CMPtwoUfour$ at $1.2\leq s\leq1.4$ and fall short of it by $\CMPtwoUfourEnds$ toward both ends. Hamiltonian truncation at finite cutoff lies above the curve by up to about 1\% at $0.175\leq\Ds\leq0.3$. The published MC values~\cite{Angelini:2014koy,Luijten:1997th,Picco:2012ak,Xiao:2025sk}, with quoted errors of $0.006$ to $0.034$, cannot separate the ways of imposing the short-range end. In $d=3$ the one accepted MC value lies $\CMPthreeMC$ above the resummed curve, while the extrapolations that are not accepted, at $1.7\leq s\leq1.9$, lie $\CMPthreeMCsr$ above it.

\section{Discussion \label{sec: discuss}}

We have studied the LRI CFT in one, two and three dimensions with five descriptions whose errors have unrelated sources. Three results stand out.

First, the perturbative data are much extended. The $\epsilon$ expansion about the generalized free field now reaches \PTorderOneDword{} loops in every dimension, with closed forms through five loops in $d=1,2$ and, in $d=3$, expressions in values of two-variable hypergeometric functions. About the short-range end, the expansion in $d=1$ reaches \SRorderword{} coefficients, the first seven in closed form and the eighth in a conjectured closed form that agrees with its numerical value to about 85 digits. In $d=2$ the leading nonvanishing short-range coefficient is now known in closed form, and in $d=3$ the estimate of the leading one is improved by computing it from the bootstrap spectrum of the short-range Ising CFT~\cite{Simmons-Duffin:2016wlq,Chang:2024whx}.

Second, Hamiltonian truncation applies to this nonlocal CFT. The GFF has a Fock space on the cylinder, and once the truncation artifact is corrected, the method reproduces the low orders of the $\epsilon$ expansion without Feynman graphs; in $d=2$, applied to the SRI CFT coupled to a GFF, it also reproduces $C^{\rm SR}_2(2)$. In $d=1$ it gives $\DE$ at finite coupling, and near the mean-field end its values agree with the $\epsilon$ expansion in every dimension.

Third, the two expansions, resummed together with the leading Borel singularity taken from an instanton, reproduce $\DE$ between the two ends. In $d=1$ the resummed curve, which uses none of the nonperturbative values, lies within $\CMPhtCB$ of every final Hamiltonian-truncation value, $\CMPnrgCB$ of every converged NRG value and $\CMPmcCB$ of every accepted MC value. Without short-range input, the eighth-order series alone stays within $\RSgffDevTwo$ of the resummed curve in $d=2$, whereas the three-loop series deviates from it by up to \RScbThreeTwo.

Universality turns these computations into tests of one another, and the test is strongest in $d=1$, where all five descriptions apply. Away from the mean-field end, Hamiltonian truncation, NRG and MC agree at the sub-percent level, and Hamiltonian truncation and NRG agree to within two parts in a thousand where the truncation is extrapolated and NRG is converged. Since the coefficients $C_n(d)$ are analytic in $d$, with graphs and renormalization common to all dimensions, these checks also test the framework behind the two- and three-dimensional series, though not their $d$-specific integrals. Hamiltonian truncation tests those integrals directly at low orders, through $C_5$ in $d=2$ and $C_4$ in $d=3$ (Sec.~\ref{sec: pert-gff}). Near the short-range end, where the truncation does not reach the short-range values, NRG reproduces the dual expansion through its fifth coefficient. This tests at these orders that our prescription renormalizes the dual correctly. The higher coefficients lie at or below the NRG resolution, so the data cannot distinguish resummed curves that impose four or eight short-range coefficients.

One cross-check is still missing: the conformal bootstrap, which uses neither a Lagrangian nor a lattice and would test all five descriptions at once. Numerical bootstrap studies of the LRI have found kinks in two and three dimensions~\cite{Behan:2018hfx,Behan:2023ile}, but the bounds of Ref.~\cite{Behan:2023ile} rest on a discretization that overestimates them, so they cannot yet be compared with our values at the precision reached here. A new bootstrap study of the LRI using analytic functionals is in progress~\cite{Ghosh:2026wip,Ghosh:2026dse,Ghosh:2023onl,Ghosh:2023wjn}.

Two families of nonlocal CFTs are natural next targets. The long-range $O(N)$ model, with $N$-component spins~\cite{Fisher:1972zz}, has an $\epsilon$ expansion with the same graphs as the LRI, each weighted by a polynomial in $N$, known to three loops~\cite{Benedetti:2020rrq}. Near its short-range end the short-range $O(N)$ CFT couples to an $N$-component GFF~\cite{Chai:2021arp}, so in $d=3$ the leading short-range coefficient would follow from the bootstrap spectrum of the $O(N)$ CFT, which is available for $N=2$~\cite{Liu:2020tpf}. Unlike the LRI, the long-range $O(N)$ model also has a $1/N$ expansion, known to next-to-leading order~\cite{Chai:2021arp}, which does not require $s$ near either end and would test resummations in the middle of the window. In the long-range minimal models, a relevant primary of the $m$-th unitary Virasoro minimal model couples to a GFF in $d=2$, and the two-dimensional LRI is the case $m=3$~\cite{Behan:2025ydd}. When the coupled primary is the Kac-table field $\phi_{(2,2)}$, the spin field for $m=3$, the mean-field end is conjectured to be described by a nonlocal $\phi^{2(m-1)}$ theory. Both ends are then weakly coupled, but both expansions are problematic at large $m$, and the intermediate range needs nonperturbative methods~\cite{Behan:2025ydd}. Several methods of this study carry over: the short-range coefficients are integrated minimal-model correlators, and the truncated cylinder Hamiltonian that reproduces $C^{\rm SR}_2(2)$ (Sec.~\ref{sec: sr}) can be built with the $m$-th minimal model in place of the Ising CFT. For $m=4$, MC simulations of a lattice model with long-range couplings at its tricritical point could supply the nonperturbative test of the conjectured duality that Ref.~\cite{Behan:2025ydd} asks for.

\section*{Note added}
We coordinated the submission of this article with that of Ref.~\cite{Antunes:2026wip} by Antunes and Matos, which uses field theory to guide Monte Carlo simulations of the one-dimensional LRI.

\section*{Data and code availability}
All code, inputs, production records and notes needed to reproduce the results of this article are released at \url{https://github.com/Canonical111/LRI_public}, which becomes public with this article; the large data files needed to rerun the calculations in full are archived at Zenodo~\cite{Zheng:2026data}. The code is under the MIT license and everything else under CC~BY~4.0, except two third-party components that keep their own licenses.

\section*{Acknowledgments}
We thank Connor Behan, Nikolay Gromov, Henry Lin and Yifan Wang for discussions and comments on the draft, and David Simmons-Duffin for access to computing time on Expanse. This work is supported by Simons Foundation grant \#994308 for the Simons Collaboration on Confinement and QCD Strings. This work used the Expanse cluster at the San Diego Supercomputer Center (SDSC) through allocation PHY190023 from the Advanced Cyberinfrastructure Coordination Ecosystem: Services \& Support (ACCESS) program, which is supported by National Science Foundation grants \#2138259, \#2138286, \#2138307, \#2137603 and \#2138296.

\pagebreak
\appendix

\section{Closed-form perturbative coefficients \label{app: closed}}

This appendix gives the closed forms behind the analytic entries of Table~\ref{tab:pt} in $d=1,2$ and the $d=2$ entry of Table~\ref{tab:sr}, coefficients of the series~\eqref{eq:eps-series} and~\eqref{eq:sr-series}; their numerical values are in those tables, and the closed forms in $d=1$ near the short-range end are Eqs.~\eqref{eq:sr-closed} and~\eqref{eq:sr-C8}. The coefficients $C_0(d)$ to $C_3(d)$, recalled below, are published~\cite{Benedetti:2020rrq}, as is the vanishing of $C^{\rm SR}_1(2)$~\cite{Behan:2017emf}; the closed forms of $C_4$ and $C_5$ in $d=1,2$ and of $C^{\rm SR}_2(2)$ are new.

\subsection{Perturbed generalized free field \label{cf:sec:gff}}

\paragraph{Through third order.} With the polygamma functions $\psi_m(z)=\mathrm{d}^{m+1}\log\Gamma(z)/\mathrm{d}z^{m+1}$ and
\begin{equation}
 A_d=\psi_0(\tfrac d2)+\psi_0(1)-2\psi_0(\tfrac d4),\qquad R_d=\psi_1(\tfrac d2)-\psi_1(1),
 \label{cf:gd-blocks}
\end{equation}
the coefficients through three loops are~\cite{Benedetti:2020rrq}
\begin{equation}
 C_0(d)=\frac d2,\qquad C_1(d)=-\frac16,\qquad C_2(d)=\frac{A_d}{9},\qquad C_3(d)=\frac{A_d^2+R_d-8Q_d}{108},
 \label{cf:gd-C03}
\end{equation}
where $Q_d$ is one third of the period of the tetrahedron graph with propagator $|x|^{-d/2}$, normalized by $h_d^2$ with $h_d=\Gamma(\frac d2)/\pi^{d/2}$, as corrected in the 2025 erratum of Ref.~\cite{Benedetti:2020rrq}. In $d=1,2,3$,
\begin{equation}
\begin{gathered}
 A_1=\pi+4\log2,\qquad R_1=\frac{\pi^2}{3},\qquad Q_1=\frac{\Gamma(\frac14)^4}{\pi},\\
 A_2=4\log2,\qquad R_2=0,\qquad Q_2=\frac{\Gamma(\frac14)^8}{24\pi^4},\\
 A_3=2+4\log2-\pi,\qquad R_3=\frac{\pi^2}{3}-4,
\end{gathered}
 \label{cf:d123-blocks}
\end{equation}
while $Q_3=4\pi^2\,\mathfrak H_3(0)-\Gamma(\frac14)^{10}/(48\pi^{7/2})$, with $\mathfrak H_3(0)$ the value of a hypergeometric function of two variables (\SM{}, Part~I, Sec.~19), which we do not reduce to functions of one variable. $C_3(3)$ is therefore of the same kind as $C_4(3)$ and $C_5(3)$.

\paragraph{Fourth and fifth order.} The expressions of $C_4(d)$ and $C_5(d)$ in general $d$, in terms of six master integrals, are given in the \SM{} (Part~I, Sec.~1.1). %
 In $d=1$ and $d=2$ the masters reduce to known constants, and the coefficients take the closed forms below.

\paragraph{One dimension.}
The periods of the four-loop wheel and of the five-loop zig-zag graph reduce to ordinary hypergeometric values at fixed rational parameters,
\begin{equation}
\begin{gathered}
 P_4(1)=13\,\Gamma(\tfrac14)^4F_--\frac{192\pi^4}{\Gamma(\frac14)^4}\,F_+,\qquad
 P_5(1)=16\pi^{3/2}\Bigl[-35\,\Upsilon_0+(22\sqrt2-13)\,\Upsilon_++(22\sqrt2+13)\,\Upsilon_-\Bigr],\\
 F_-={}_4F_3\!\left(\begin{matrix}\frac14,\frac14,\frac12,\frac12\\[2pt]\frac34,\frac34,1\end{matrix};1\right),\qquad
 F_+={}_4F_3\!\left(\begin{matrix}\frac34,\frac34,\frac12,\frac12\\[2pt]\frac54,\frac54,1\end{matrix};1\right),\\
 \Upsilon_0=\mathcal Y_2,\qquad \Upsilon_+=c^7\mathcal Y_3-s^7\mathcal Y_5,\qquad \Upsilon_-=c^7\mathcal Y_1+s^7\mathcal Y_7,\qquad c=\cos\tfrac\pi8,\quad s=\sin\tfrac\pi8,\\
 \mathcal Y_r=\frac r8\left[\frac{\Gamma\big(\frac{r+2}8\big)}{\Gamma\big(\frac{r+6}8\big)}\right]^7
 {}_9F_8\!\left(\begin{matrix}1,\;1+\frac r8,\;\big(\frac{r+2}8\big)^{\times7}\\[2pt]\frac r8,\;\big(\frac{r+6}8\big)^{\times7}\end{matrix};-1\right),\qquad r=1,2,3,5,7,
\end{gathered}
\label{cf:d1periods}
\end{equation}
with $a^{\times k}$ a parameter repeated $k$ times and
\begin{equation}
 {}_pF_q\!\left(\begin{matrix}a_1,\ldots,a_p\\ b_1,\ldots,b_q\end{matrix};x\right)
 =\sum_{n=0}^{\infty}\frac{(a_1)_n\cdots(a_p)_n}{(b_1)_n\cdots(b_q)_n}\,\frac{x^n}{n!},
 \qquad (a)_n=\frac{\Gamma(a+n)}{\Gamma(a)} .
 \label{cf:pFq}
\end{equation}
With Eq.~\eqref{cf:d123-blocks} and Catalan's constant $\mathrm G$,
\begin{align}
 648\,C_4(1)&=16P_4(1)-A_1^3-\pi^2A_1-53\pi^3-1436\,\zeta(3)-8A_1Q_1 ,
 \label{cf:d1C4}\\
 46656\,C_5(1)&=-1536\,P_5(1)+96(7A_1-\pi)P_4(1)+384\pi\,\Gamma(\tfrac14)^4F_--1728Q_1^2\notag\\
 &\quad+16Q_1\bigl(45A_1^2+319\pi^2+3576\,\mathrm G\bigr)-196608\,\mathrm G^2-36864\pi^2\mathrm G\notag\\
 &\quad-15A_1^4-30\pi^2A_1^2+708\pi^3A_1+21552A_1\,\zeta(3)-1619\pi^4 .
 \label{cf:d1C5}
\end{align}
The reductions of the masters behind these formulas are proved in the \SM{}, so $C_4(1)$ and $C_5(1)$ are closed forms.

\paragraph{Two dimensions.}
With Eq.~\eqref{cf:d123-blocks} and the one-dimensional values~\eqref{cf:d1periods},
\begin{align}
 648\,C_4(2)&=16P_4(2)-64\log^32-32\log2\,Q_2-359\,\zeta(3),
 \label{cf:d2-C04}\\
 972\,C_5(2)&=-32P_5(2)+56\log2\,P_4(2)-36Q_2^2+\bigl(240\log^22+596\,\mathrm G\bigr)Q_2\notag\\
 &\quad-1024\,\mathrm G^2+449\,\zeta(3)\log2-80\log^42 ,
 \label{cf:d2-C5}
\end{align}
where the two-dimensional periods are quadratic in the one-dimensional values,
\begin{equation}
\begin{gathered}
 P_4(2)=\frac1{4\pi^3}\left[\Gamma(\tfrac14)^8F_-^2-\frac{4096\pi^8}{\Gamma(\frac14)^8}F_+^2\right],\\
 P_5(2)=\frac8\pi\Bigl[25\Upsilon_0^2-28\sqrt2\,\Upsilon_0(\Upsilon_++\Upsilon_-)+18\bigl(\Upsilon_+^2+\Upsilon_-^2\bigr)+28\Upsilon_+\Upsilon_-\Bigr].
\end{gathered}
 \label{cf:d2-P}
\end{equation}
The reductions of the masters are proved and the factorizations~\eqref{cf:d2-P} are derived analytically, with no integer-relation search; both are given in the \SM{}.

\subsection{Short-range end in two dimensions \label{cf:sec:cr2}}

In $d=2$, where $s_*=7/4$, the coefficient $C^{\rm SR}_2(2)$ of Eq.~\eqref{eq:sr-series} is a closed form in one Gamma-function ratio,
\begin{equation}
 R_\Gamma=\frac{\Gamma(1/4)}{\Gamma(3/4)},
 \label{cf:cr-RGamma}
\end{equation}
and three hypergeometric values, in the convention~\eqref{cf:pFq},
\begin{equation}
 \mathsf P={}_4F_3\!\left(\begin{matrix}-\frac12,-\frac12,\frac32,\frac34\\[2pt]1,2,\frac14\end{matrix};1\right),\qquad
 \mathsf Q={}_4F_3\!\left(\begin{matrix}\frac12,\frac32,\frac32,\frac94\\[2pt]3,3,\frac74\end{matrix};1\right),\qquad
 \mathsf T={}_4F_3\!\left(\begin{matrix}\frac32,-\frac14,\frac34,\frac34\\[2pt]\frac14,\frac94,\frac94\end{matrix};1\right).
 \label{cf:cr-PQT}
\end{equation}
In terms of these,
\begin{equation}
 C^{\rm SR}_2(2)=\frac{16}{R_\Gamma^2}-\frac{R_\Gamma^2}{9}
 -\frac{\pi^2R_\Gamma^2}{16}
 \left(\mathsf P-\frac{5R_\Gamma^2}{384}\mathsf Q\right)
 \left(\mathsf P-\frac{5R_\Gamma^2}{128}\mathsf Q
             -\frac{512}{25\pi R_\Gamma^2}\mathsf T\right).
 \label{cf:cr-c2}
\end{equation}
In each of the series~\eqref{cf:cr-PQT}, $\sum_jb_j-\sum_ia_i=2$, so the terms decay as $n^{-3}$ and the series converge absolutely.

Equation~\eqref{cf:cr-c2} is a closed form in named special functions; no reduction to Gamma values alone has been established. The identification with the physical coefficient is a formal perturbative result in a critically tuned analytic minimal-subtraction prescription. A bound on the remainder of the nested subtractions, uniform in the regulator, is not proved. The derivation starts from the exact correlators of the two-dimensional Ising CFT~\cite{Ardonne:2010hj} and reduces the integrals over the plane to bilinear combinations of hypergeometric periods, as in Ref.~\cite{Duhr:2023bku}; it is given in the \SM{}.

\section{The truncation artifact and its correction \label{app: window}}

This appendix writes out the truncated Hamiltonian of Sec.~\ref{sec: model-ht} in $d=1$ and uses it to explain the artifact of Hamiltonian truncation and how we correct it. The \SM{} gives the construction in every dimension (Part~IV, Sec.~3) and its analysis order by order in $\epsilon$ (Part~III, Sec.~2.7).

\textit{Hamiltonian.} In $d=1$ the sphere $S^0$ is the pair of points $x=\pm1$, and the GFF on the cylinder is
\begin{equation}
 \phi(\pm1)=\sum_{n\geq0}u_n\,(\pm1)^n\big(a_n+a_n^\dagger\big),\qquad u_n^2=\frac{(2\Ds)_n}{n!},
 \label{eq:win-modes}
\end{equation}
with one oscillator $a_n$ of energy $\Ds+n$ for each $n\geq0$. A Fock state with $P$ quanta and level $\ell=\sum_n n\,a_n^\dagger a_n$ has free energy $P\Ds+\ell$. The reflection $x\to-x$ acts as $a_n\to(-1)^na_n$, and within each reflection sector the average over $S^0$ in Eq.~\eqref{eq:HT} equals the value at $x=+1$. There we split the field into the zero mode and the nonzero modes,
\begin{equation}
 \phi(+1)=q+\chi,\qquad q=a_0+a_0^\dagger,\quad p=\ii\big(a_0^\dagger-a_0\big),\qquad \chi=\chi_-+\chi_-^\dagger,\quad \chi_-=\sum_{n\geq1}u_n\,a_n .
 \label{eq:win-split}
\end{equation}
Normal ordering acts on all oscillators, so ${:}q^2{:}=q^2-1$, ${:}q^3{:}=q^3-3q$ and ${:}q^4{:}=q^4-6q^2+3$. Up to a constant, the Hamiltonian~\eqref{eq:HT} is then
\begin{equation}
\begin{gathered}
 H=H_z\otimes1+1\otimes H_\chi+\sum_{j=1}^4L_j\otimes W_j,\qquad W_j={:}\chi^j{:}=\sum_{k=0}^j\binom jk\big(\chi_-^\dagger\big)^k\chi_-^{\,j-k},\\
 H_z=\frac{\Ds}4\,p^2+A\,q^2+\frac g{12}\,q^4,\qquad A=\frac{\Ds}4+r-\frac g2,\qquad H_\chi=\sum_{n\geq1}(\Ds+n)\,a_n^\dagger a_n,\\
 L_1=(2r-g)\,q+\frac g3\,q^3,\qquad L_2=r-\frac g2+\frac g2\,q^2,\qquad L_3=\frac g3\,q,\qquad L_4=\frac g{12}.
\end{gathered}
\label{eq:win-H}
\end{equation}
$H_z$ is the zero mode with the part of the interaction that acts on it alone; at the tuned couplings $A$ can be negative, and $H_z$ is then a double well. The $L_j$ couple the zero mode to the nonzero modes. The split is needed at finite coupling, because the zero mode carries the states we tune and predict: at weak coupling $\sigma$, $\mathcal E$ and $\phi^3$ are its excitations $(a_0^\dagger)^p|0\rangle$ with $p=1$, 2 and 3. Its quanta add no level and cost only the energy $\Ds$, which vanishes toward $s=1$, while the tuned $g$ grows with $N$, so the fluctuations of the zero mode are large. A cutoff on the number of its quanta, such as the energy cutoff that suffices order by order in $\epsilon$, where each order involves a bounded number of quanta, would truncate these states. We therefore do not cap the number of zero-mode quanta. We diagonalize the full Hamiltonian~\eqref{eq:win-H}, the couplings $L_j\otimes W_j$ included, on the product of a zero-mode basis with the Fock states of the nonzero modes of level $\ell\leq N$, with no limit on the number of their quanta. The zero mode is truncated too, to \HTnpDensK\ states per parity, but each of them is a superposition of many quanta. These states are the leading eigenvectors of the zero-mode density matrix of the coupled low-lying states, and they reproduce the $\DE$ of a converged zero mode to $\HTnpBasisErr$ (\SM{}, Part~IV, Sec.~2.5). A basis of the lowest eigenstates of $H_z$ alone converges more slowly in the number of states, because the coupling to the nonzero modes changes the zero-mode state, the more so as $g$ grows with $N$. The normal-ordered $W_j$ never pass through a level above both of their endpoints, so the Hamiltonian $H(M)$ at a cutoff $M<N$ is exactly the projection of $H(N)$ onto $\ell\leq M$. The symmetry $\phi\to-\phi$ splits each reflection sector by the parity of $P$. In the reflection-even sector, the two lowest states of even $P$ are the vacuum and $\mathcal E$, and the two lowest of odd $P$ are $\sigma$ and $\phi^3$. At each $N$ the couplings $r$ and $g$ are tuned so that $\sigma$ and $\phi^3$ lie $\Ds$ and $1-\Ds$ above the vacuum energy $E_v$, and $\DE$ is the gap of $\mathcal E$.

\textit{Vacuum bubbles.} Acting on a state, the interaction can create quanta that do not touch the quanta already present, and a later action can annihilate them again. These disconnected processes, the vacuum bubbles, are the vacuum fluctuations of an excited state. In the full theory every state carries the same bubbles as the vacuum, so they shift all energies alike and cancel in the gaps $E-E_v$. With a cutoff, a bubble of level $\ell_b$ fits on a state of level $\ell$ only if $\ell+\ell_b\leq N$. The state keeps the bubbles up to level $N-\ell$ and the vacuum keeps them up to $N$, so the state misses those in the range $N-\ell<\ell_b\leq N$ and its gap comes out too high.

At second order in $g$ the missing amount is explicit. The bubbles of $V=\frac{g}{12}\langle{:}\phi^4{:}\rangle$ are four-quantum states $|b\rangle$ of level $\ell_b$ and energy $4\Ds+\ell_b$, and a state of level $\ell$ lies too high by
\begin{equation}
 -D(\ell)=\sum_{b:\,N-\ell<\ell_b\leq N}\frac{|\langle b|V|0\rangle|^2}{4\Ds+\ell_b},\qquad D(\ell)=E_v(N)-E_v(N-\ell),
 \label{eq:win-second}
\end{equation}
where $E_v(M)$ is the vacuum energy at cutoff $M$. This is the counterpart, with a level cutoff, of the state-dependent counterterm that Ref.~\cite{Elias-Miro:2020qwz} adds with an energy cutoff in $\phi^4$ theory in $2+1$ dimensions. Since $\sum_nu_n^2x^n=(1-x)^{-2\Ds}$, the bubbles of level $\ell_b$ carry the total weight $24(g/12)^2$ times the coefficient of $x^{\ell_b}$ in $(1-x)^{-8\Ds}$; the average over $S^0$ keeps only even $\ell_b$. At the mean-field end $\Ds=\frac14$, where ${:}\phi^4{:}$ has dimension 1 and is marginal, this coefficient is $\ell_b+1$, equal to the energy. Every even level in this range then contributes exactly $g^2/6$, whatever $N$ is. With $g\simeq\epsilon/3$ near that end, a level-1 descendant lies too high by $\epsilon^2/54$ at leading order for every even $N$. The states of $\sigma$, $\phi^3$ and $\mathcal E$ have level 0 at this order and miss nothing.

\textit{At finite coupling.} Conformal symmetry fixes the descendant spacings, $E(\partial^n\mathcal O)-E(\mathcal O)=n$, so their violations $v_n$ measure the artifact with no continuum input. We identify $\partial\mathcal O$ by its overlap with $L_{-1}\mathcal O$, with $L_{-1}$ the generator of the free GFF. Without the correction, the level-1 violations at $\Ds\leq0.1$ are $\HTnpRawViol$ for $N=\HTnpViolN$, and they do not shrink as $N$ grows; Table~\ref{tab:win} gives them at $N=\HTnpIdentN$.

\begin{table}[H]
\centering
\renewcommand{\arraystretch}{1.15}
\small
\begin{tabular}{@{}l r r r r r r@{}}
\toprule
 & \multicolumn{3}{c}{uncorrected} & \multicolumn{3}{c}{corrected}\\
\cmidrule(lr){2-4}\cmidrule(l){5-7}
$\Ds$ & $\sigma$ & $\phi^3$ & $\mathcal E$ & $\sigma$ & $\phi^3$ & $\mathcal E$\\
\midrule
0.2 & $+0.0030$ & $+0.0067$ & $+0.0048$ & $+0.0007$ & $+0.0044$ & $+0.0024$\\
0.15 & $+0.0207$ & $+0.0297$ & $+0.0238$ & $+0.0007$ & $+0.0096$ & $+0.0040$\\
0.1 & $+0.0644$ & $+0.0743$ & $+0.0580$ & $-0.0044$ & $+0.0046$ & $-0.0100$\\
0.075 & $+0.0990$ & $+0.1056$ & $+0.0797$ & $-0.0088$ & $-0.0024$ & $-0.0248$\\
0.05 & $+0.1481$ & $+0.1439$ & $+0.1040$ & $-0.0123$ & $-0.0110$ & $-0.0423$\\
0.025 & $+0.2321^a$ & $+0.1869$ & $+0.1273$ & $-0.0063^a$ & $-0.0186$ & $-0.0548$\\
\bottomrule
\end{tabular}
\caption{Violation $v_1=E(\partial\mathcal O)-E(\mathcal O)-1$ of the level-1 descendant spacing in $d=1$ at $N=\HTnpIdentN$, without and with the correction of Eq.~\eqref{eq:win-shift}. In the CFT every entry vanishes. For $\mathcal E$, $\mathcal O$ is the first excited even state. $^a$:~the identification of $\partial\mathcal O$ is ambiguous.}
\label{tab:win}
\end{table}

\textit{The correction.} The correction is an additional term in the Hamiltonian. At cutoff $N$ we diagonalize
\begin{equation}
 H_{\rm eff}(N)=H(N)+1\otimes D_N(\hat\ell),\qquad \hat\ell=\sum_{n\geq1}n\,a_n^\dagger a_n,\qquad D_M(\ell)=E_v(M)-E_v(M-\ell),
 \label{eq:win-shift}
\end{equation}
where $H(M)$ is the Hamiltonian~\eqref{eq:win-H} at cutoff $M$ and $E_v(M)$ is the lowest eigenvalue of $H_{\rm eff}(M)$. The added term is diagonal in the Fock basis of the nonzero modes: it lowers every state of level $\ell$, in every sector, by $-D_N(\ell)$, and it leaves the states of level 0 untouched. It is not a local operator but a function of the level, fixed by the vacuum energies of the corrected Hamiltonian at the lower cutoffs. These vacuum energies include the zero mode, whose quanta enter the bubbles at no cost in level; vacua of the nonzero modes alone remove only part of the violations. No parameter is fitted. Reference~\cite{Elias-Miro:2020qwz} adds such terms order by order, including towers of repeated bubbles; Eq.~\eqref{eq:win-shift} instead takes all disconnected bubbles at once from the vacuum energies at lower cutoffs. At second order $D_N$ reduces to $D$ of Eq.~\eqref{eq:win-second}. At $\Ds=\frac14$ it is then $-g^2/6$ times the number of even levels in $(N-\hat\ell,N]$, which is $\hat\ell/2$ up to parity, so the added term is close to $-\frac{g^2}{12}\hat\ell$. It lowers the coefficient of the level in $H_\chi=\hat\ell+\Ds\sum_{n\geq1}a_n^\dagger a_n$ from 1 to $1-g^2/12$ and removes the violations of the descendant spacings at this order.

\textit{Tuning with the correction.} Each cutoff $N$ is tuned separately: the two tuning conditions are two equations for the two unknowns $r$ and $g$, solved by Newton's method. One iteration, at a trial pair $(r,g)$, has five steps.
\begin{enumerate}
\item \emph{Vacuum energies at lower cutoffs.} Compute $E_v(M)$ for $M=0,2,\dots,N$ in increasing order, at the current $(r,g)$, with the same $\Ds$ and the same zero-mode basis; no lower cutoff is retuned. $E_v(0)$ is the lowest eigenvalue of $H(0)$, which acts on the zero mode alone, and $E_v(M)=E_v(M-1)$ for odd $M$, since the reflection-even sector, which contains the vacuum, has only even levels.
\item \emph{Fixed point.} Since $D_M$ contains $E_v(M)$ itself, each $E_v(M)$ solves the fixed-point equation $E_v(M)=\lambda_{\min}\big[H(M)+1\otimes D_M(\hat\ell)\big]$ in one variable, which an inner Newton iteration solves.
\item \emph{Diagonalization.} Diagonalize $H_{\rm eff}(N)$. Its lowest eigenvalue is $E_v(N)$, and the gaps are measured from it.
\item \emph{Residuals.} Compute $F_1=E_\sigma-E_v-\Ds$ and $F_2=E_{\phi^3}-E_v-(1-\Ds)$. Both vanish when the tuning conditions hold.
\item \emph{Newton step.} Take a Newton step toward $F_1=F_2=0$ in coordinates $\theta=(\theta_0,\theta_1)$ that parametrize $(r,g)$ (\SM{}, Part~IV, Sec.~2.6). The direction is $-J^{-1}(F_1,F_2)$, with $J$ the matrix of derivatives of $(F_1,F_2)$ with respect to $\theta$, and the step is halved until $|(F_1,F_2)|$ decreases. Since $D_N$ contains the $E_v(M)$, which depend on the couplings, $J$ includes the response of every $E_v(M)$, obtained from the Hellmann--Feynman theorem. Return to step~1 until $F_1$ and $F_2$ vanish within a fixed tolerance.
\end{enumerate}
At the solution, $\DE=E_{\mathcal E}-E_v$ is the prediction. The $E_v(M)$ with $M<N$ that enter $H_{\rm eff}(N)$ are vacuum energies of the theory at the couplings tuned for $N$, truncated at level $M$, not those of the theory tuned at cutoff $M$.

\textit{What it removes.} At $N=\HTnpIdentN$ and $\Ds\leq0.15$ the correction reduces the violation for $\sigma$ by a factor of $\HTnpWinReduce$. It restores only the disconnected bubbles, so at finite coupling it does not remove the artifact completely. At $\Ds\leq0.1$ residual violations that do not shrink with $N$ remain: about $\HTnpResSig$ for $\sigma$ and $\phi^3$, and up to $\HTnpResE$ for $\mathcal E$ at $\Ds\leq0.05$. At $\Ds\leq0.025$ the corrected sequence of $\DE$ has no limit as $N$ grows, a property of the correction rather than of the numerics; the uncorrected truncation converges there, but to values that violate the descendant spacings. Expanded in $\epsilon$, the correction changes $\DE$ only at $O(\epsilon^4)$: it does not act on states of level 0, and its shifts of order $g^2$ enter their energies through intermediate states at order $g^4$. It therefore leaves $C_2$ and $C_3$ unchanged. At $\Ds=0.075$, 0.1 and 0.125 the same extrapolations without the correction give values $\HTnpRawShift$ lower (Sec.~\ref{sec: compare}).

\section{Numerical results \label{app: numerics}}

Tables~\ref{tab:num1} to~\ref{tab:num3} give, for $d=1,2,3$, the best value of $\DE$ from each description at each coupling, as shown in Figs.~\ref{fig:cmp1d} and~\ref{fig:cmp2d3d}; $s=d-2\Ds$. The research notes in the \SM{} give the full results, with their error budgets and the status of each fit. Only the MC statistical errors and the errors of the effective MC exponents are standard deviations. The resummed values identify the approximant; their digits are not claimed digits of $\DE$.

\begin{table}[H]
\centering
\renewcommand{\arraystretch}{1.15}
\small
\begin{tabular}{@{}l l l l l l@{}}
\toprule
$\Ds$ & $s$ & HT & NRG & MC & Resummed\\
\midrule
0.001 & 0.998 & $[\HTnpNfortyOne]$ & $\NRGDEOne$ & & 0.93742\\
0.01 & 0.98 & $[\HTnpNfortyTen]$ & $\NRGDETen$ & & 0.80975\\
0.0125 & 0.975 & & & $[0.78404(77)(1003)]$ & 0.78915\\
0.025 & 0.95 & $[\HTnpNfortyTwentyFive]$ & $\NRGDETwentyFive$ & $0.71115(49)(959)$ & 0.71343\\
0.0375 & 0.925 & & & $[0.65904(51)(425)]$ & 0.66146\\
0.05 & 0.9 & $[\HTnpNfortyFifty]$ & $\NRGDEFifty$ & $0.61987(55)(220)$ & 0.62223\\
0.0625 & 0.875 & & & $[0.59003(35)(150)]$ & 0.59139\\
0.075 & 0.85 & $\HTnpDEseventyFive$ & $\NRGDESeventyFive$ & $0.56694(37)(139)$ & 0.56663\\
0.0875 & 0.825 & & & $0.54761(32)(194)$ & 0.54656\\
0.1 & 0.8 & $\HTnpDEoneHundred$ & $\NRGDEOneHundred$ & $0.53248(20)(242)$ & 0.53025\\
0.1125 & 0.775 & & & $0.52020(25)(217)$ & 0.51703\\
0.125 & 0.75 & $\HTnpDEoneTwentyFive$ & $\NRGDEOneTwentyFive$ & $0.50710(35)(193)$ & 0.50643\\
0.1375 & 0.725 & & & $0.49856(30)(180)$ & 0.49808\\
0.15 & 0.7 & $\HTnpDEoneFifty$ & $\NRGDEOneFifty$ & $[0.49325(41)(306)]$ & 0.49171\\
0.1625 & 0.675 & & & $[0.48761(58)]^{\rm P}$ & 0.48714\\
0.1727335 & 0.6545 & & & $0.48368(52)(78)^{\rm P}$ & 0.48465\\
0.175 & 0.65 & $\HTnpDEoneSeventyFive$ & $\NRGDEOneSeventyFive$ & $[0.48460(35)(67)]$ & 0.48424\\
0.1875 & 0.625 & & & $[0.48263(30)]$ & 0.48291\\
0.2 & 0.6 & $\HTnpDEtwoHundred$ & $\NRGDETwoHundred$ & & 0.48312\\
0.225 & 0.55 & $\HTnpDEtwoTwentyFive$ & $\NRGDETwoTwentyFive$ & $[0.47912(68)]$ & 0.48808\\
0.24 & 0.52 & $\HTnpDEtwoForty$ & $\NRGDETwoForty$ & & 0.49422\\
0.249 & 0.502 & $\HTnpDEtwoFortyNine$ & $\NRGDETwoFortyNine$ & & 0.49934\\
\bottomrule
\end{tabular}
\caption{$\DE$ in $d=1$. Brackets: not extrapolated, not converged, or not accepted (open symbols in Fig.~\ref{fig:cmp1d}). HT, Hamiltonian truncation: the limit $N\to\infty$ with half the spread of the accepted fits, probably optimistic at $\Ds\geq0.15$, where we estimate the total error at about $\HTnpTrust$; at 0.075 a second, one-sided error; in brackets, the value at $N=\HTnpNmax$; at 0.05 the fits give $\HTnpDEfiftyFOneTwo$, or down to $\HTnpDEfiftyFthree$. NRG: a linear sum of chain-length, truncation and discretization errors; in brackets, not converged in chain length. MC: the value with its statistical error and fit-model half-range, both in units of $10^{-5}$; in brackets, not accepted by the fit protocol; P, the status depends on the computing platform; at $\Ds=0.2$, 0.2125 and 0.2375 no fit passes. Resummed: the curve of Sec.~\ref{sec: resum}; prescriptions spread over at most $\RSoneDenv$.}
\label{tab:num1}
\end{table}

\begin{table}[H]
\centering
\renewcommand{\arraystretch}{1.15}
\small
\begin{tabular}{@{}l l l l l@{}}
\toprule
$\Ds$ & $s$ & HT, $N=\HTnpDtwoNmax$ & MC, $L=512\to1024$ & Resummed\\
\midrule
0.125 & 1.75 & $[0.94003]$ & & 1\\
0.126 & 1.748 & $[0.94568]$ & & 1.00000\\
0.135 & 1.73 & 0.96203 & & 0.99962\\
0.15 & 1.7 & 0.99023 & $0.9773(17)$ & 0.99780\\
0.175 & 1.65 & 0.99972 & & 0.99242\\
0.2 & 1.6 & 0.99530 & $0.9698(18)$ & 0.98542\\
0.25 & 1.5 & 0.97771 & $0.9661(18)$ & 0.97114\\
0.3 & 1.4 & 0.96349 & $0.9627(20)$ & 0.96118\\
0.35 & 1.3 & 0.95782 & $0.9584(20)$ & 0.95810\\
0.4 & 1.2 & 0.96179 & $0.9631(22)$ & 0.96257\\
0.45 & 1.1 & 0.97528 & $0.9659(23)$ & 0.97543\\
0.475 & 1.05 & 0.98581 & $0.9702(25)$ & 0.98584\\
0.495 & 1.01 & 0.99678 & & 0.99678\\
\bottomrule
\end{tabular}
\caption{$\DE$ in $d=2$. HT: the truncated model at its largest cutoff, not extrapolated; in brackets, $N=22$, since at $N=\HTnpDtwoNmax$ the tuning conditions have no solution there. MC: no infinite-size value is accepted; the effective exponent of the Binder cumulant from the two largest lattices, with its bootstrap error. Resummed: the curve of Sec.~\ref{sec: resum}, with the value $\DE=1$, the vanishing slope and the curvature $C^{\rm SR}_2(2)$ imposed at $\Ds=\frac18$; prescriptions spread over at most $\RStwoDenv$, and the way the end is imposed moves the curve by up to $\RStwoDendpoint$.}
\label{tab:num2}
\end{table}

\begin{table}[H]
\centering
\renewcommand{\arraystretch}{1.15}
\small
\begin{tabular}{@{}l l l l l@{}}
\toprule
$\Ds$ & $s$ & HT, $N=\HTnpDthreeNmax$ & MC & Resummed\\
\midrule
0.518149 & 1.9637 & $\HTnpDthreeSR$ & & 1.41263\\
0.52 & 1.96 & 1.41951 & & 1.41313\\
0.525 & 1.95 & 1.42026 & & 1.41449\\
0.535 & 1.93 & 1.42182 & & 1.41719\\
0.55 & 1.9 & 1.42435 & $[1.45449(720)(452)]$ & 1.42123\\
0.575 & 1.85 & 1.42923 & $[1.45830(103)(9064)]$ & 1.42807\\
0.6 & 1.8 & 1.43511 & & 1.43524\\
0.625 & 1.75 & 1.44218 & & 1.44299\\
0.65 & 1.7 & 1.45062 & $[1.47375(174)(5237)]$ & 1.45154\\
0.675 & 1.65 & 1.46053 & & 1.46116\\
0.7 & 1.6 & 1.47192 & $\MCthreeD$ & 1.47214\\
0.725 & 1.55 & 1.48486 & $[1.48118(1632)(2716)]$ & 1.48488\\
0.74 & 1.52 & 1.49360 & & 1.49360\\
0.745 & 1.51 & 1.49674 & & 1.49674\\
0.7499 & 1.5002 & 1.49993 & & 1.49993\\
\bottomrule
\end{tabular}
\caption{$\DE$ in $d=3$. HT: the truncated model at its largest cutoff, not extrapolated. MC: the value with its statistical error and fit-model half-range, both in units of $10^{-5}$; the value at $\Ds=0.7$ is accepted only under an amendment of the protocol written after the evaluations; in brackets, not accepted; at $\Ds=0.525$, 0.6 and 0.625 no fit passes. Resummed: the curve of Sec.~\ref{sec: resum}, with the short-range value and the slope $C^{\rm SR}_1(3)$ imposed at $\Ds=0.518149$; prescriptions spread over at most $\RSthreeDenv$.}
\label{tab:num3}
\end{table}

\bibliography{ref}
\bibliographystyle{JHEP}

\end{document}